\documentclass[prd,showpacs,amsmath,amssymb,eqsecnum,nofootinbib,pdflatex]{revtex4-2}
\usepackage{float}
\usepackage{dcolumn}
\usepackage{amsfonts,mathrsfs}
\usepackage{bm}
\usepackage{subfigure}
\usepackage{url}
\usepackage{hyperref,graphicx,color}
\newtheorem{Def}{Definition}
\begin{document}
\preprint{RUP-21-20}
\preprint{KEK-Cosmo-0278}
\preprint{KEK-TH-2353}
\title{
Complete classification of 
Friedmann-Lema\^{i}tre-Robertson-Walker solutions 
with 
linear equation of state:
parallelly propagated curvature singularities for general geodesics}
%
\author{Tomohiro Harada}
\email{harada@rikkyo.ac.jp}
\affiliation{Department of Physics, Rikkyo University, Toshima,
Tokyo 171-8501, Japan}
\author{Takahisa Igata}
\email{takahisa.igata@gakushuin.ac.jp}
\affiliation{KEK Theory Center, Institute of Particle and Nuclear Studies,
High Energy Accelerator Research Organization, Tsukuba 305-0801, Japan}
\affiliation{
Department of Physics, Gakushuin University, Mejiro, Toshima-ku, Tokyo 
171-8588, Japan
}
\author{Takuma Sato}
\email{stakuma@rikkyo.ac.jp}
\affiliation{Department of Physics, Rikkyo University, Toshima,
Tokyo 171-8501, Japan}
\author{Bernard 
Carr}
\email{b.j.carr@qmul.ac.uk}
\affiliation{School of Physics and Astronomy, Queen Mary University of London, Mile End Road, London E1 4NS, United Kingdom}
\date{\today}
\begin{abstract}
We completely classify the Friedmann-Lema\^{i}tre-Robertson-Walker solutions with 
spatial curvature $K=0,\pm 1$ for perfect fluids with linear equation of state $p=w\rho $,  where $\rho$ and $p$ are the energy density and 
pressure, without assuming any energy conditions.
We extend our previous work to include all geodesics and parallelly 
propagated curvature singularities, 
showing that no non-null geodesic 
emanates from or 
terminates at the null portion of 
conformal infinity 
and that the initial singularity for $K=0,-1$ and $-5/3<w<-1$ is a null non-scalar polynomial curvature singularity.  We thus obtain the Penrose diagrams for all possible cases and  identify $w=-5/3$ as a critical value for
both the future big-rip singularity 
and the past null conformal boundary.
\end{abstract}

\maketitle



\section{Introduction and summary}
The Friedmann-Lema\^{i}tre-Robertson-Walker (FLRW) spacetime is unique as a spatially homogeneous and 
isotropic spacetime~\cite{Wald:1984rg}.
Its metric has spatial curvature $K=0,\pm 1$ and it is 
generally accepted that this, together with the Einstein equation, 
approximately describes our Universe.
For the matter content, a perfect fluid with 
linear equation of state $p=w\rho $ is often adopted, where 
$\rho$ and $p$ are the energy density and pressure,
respectively,  and $w$ is a constant.
Recent
observations strongly suggest the acceleration of the cosmological
expansion. This implies the existence of
dark energy with
equation of state parameter 
$w<-1/3$.  Cosmological observations require 
$w=-1.00^{+0.04}_{-0.05}$~\cite{DES:2017myr} and restrict the
spatial curvature to $\Omega_{K}=0.000\pm 0.005$~\cite{Planck:2015fie}.

Linear equations of state with $w=1/3$, $0$ and $-1$ correspond to a Universe dominated by radiation,
dust and a cosmological constant,
respectively.
Phenomenologically, $w=1$, $-1/3$ and $-2/3$ correspond to a massless scalar field (or stiff fluid), a string network and a domain wall network, respectively.
The ranges $-1/3<w<-1$ and $w<-1$ correspond 
to quintessence~\cite{Caldwell:1997ii,Zlatev:1998tr} and a phantom 
field~\cite{Caldwell:1999ew}, respectively. 
The null energy condition corresponds to $(1+w)\rho \ge 0$  but
viable modified theories of gravity may 
violate this~\cite{Kobayashi:2019hrl}.
The FLRW spacetime not only provides a realistic model 
for the {\it whole} Universe; it may also represent {\it part} of the Universe.
The most famous example is the Oppenheimer-Snyder solution~\cite{Oppenheimer:1939ue},
where the FLRW solution describes the metric inside a uniform ball which is collapsing
to a black hole.
FLRW spacetimes can also be used to
model 
the formation of a primordial black hole~\cite{Carr:1975qj,Harada:2013epa} 
and a stable 
gravastar~\cite{Mazur:2004fk,Visser:2003ge}.

The conformal structure and singularities are the basic properties of any spacetime.
The Penrose diagrams for the flat FLRW solutions are 
summarised in reference~\cite{Senovilla:1998oua} for models satisfying 
the dominant energy condition ($\rho\ge 0$ and $-1\le w\le 1$). 
The possibility of big-rip singularities has been proposed for $w<-1$ in references~\cite{Caldwell:2003vq,Dabrowski:2003jm,Dabrowski:2004hx} and
the behaviour of geodesics around the big-bang, big-crunch and big-rip singularities
has been analysed in detail in reference~\cite{Fernandez-Jambrina:2006tkb}.
More exotic types of FLRW singularities, such as weak 
ones,
are investigated in 
reference~\cite{Dabrowski:2018ucy}.
Recently, the past extendibility of inflationary FLRW Universes has been discussed 
in references~\cite{Yoshida:2018ndv,Nomura:2021lzz}.

In reference~\cite{Harada:2018ikn}, which we will refer to as Paper I, 
all of the FLRW solutions with $p=w\rho $ are given without assuming 
any energy conditions and 
 the corresponding Penrose diagrams are
drawn, analysing
null and comoving timelike geodesics, 
scalar polynomial (s.p.) curvature singularities,
trapped regions and trapping horizons.
This paper extends that analysis to include all geodesics 
and parallelly propagated (p.p.) curvature singularities.
This enables us to study the physical properties of singularities, such as the fate of 
free-falling observers,  in complete generality. 
We 
conclude that the null portion of conformal infinity 
repels non-null geodesics in the Penrose diagram,  so that no non-null geodesic can emanate from or terminate at the null portion of conformal infinity.
By introducing p.p. curvature singularities, which do not necessarily involve the divergence of 
any fluid quantities, we can 
study the extendibility of the spacetime beyond the conformal boundaries with 
$C^{2-}$ metrics.
We conclude that the flat and negative-curvature solutions are past inextendible beyond the null boundary for $-5/3<w<-1$, which means that these cases also have an 
initial singularity problem.

\begin{figure}[H]
\begin{center}
\includegraphics[width=0.2\textwidth]{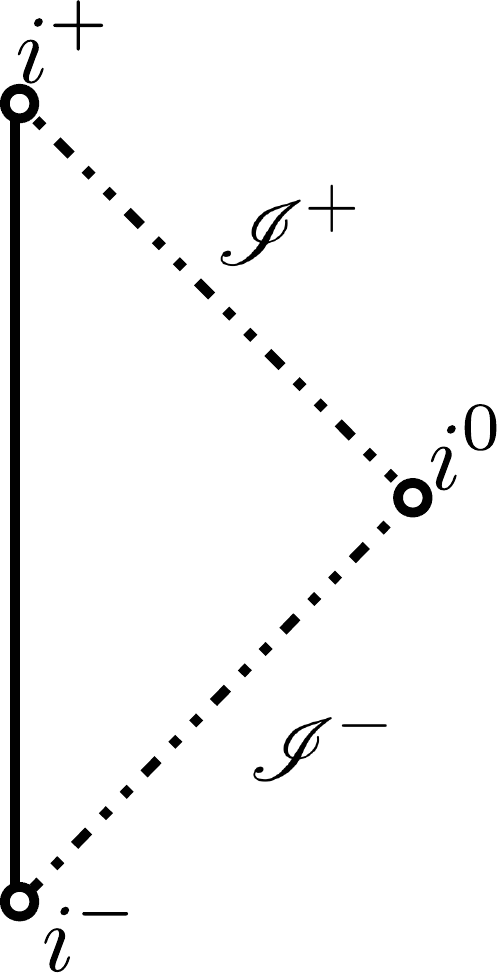}
\end{center}
\caption{The Penrose diagram for Minkowski spacetime, included
 for completeness. \label{fg:Minkowski_penrose}}
\end{figure}

\begin{figure}[H]
\begin{center}
 \begin{tabular}{cccc}
\subfigure[$w>-1/3$ (F1)]{\includegraphics[width=0.2\textwidth]{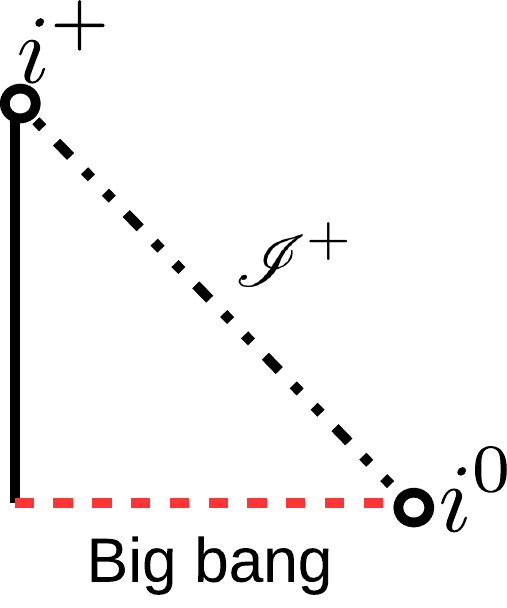}}
  &
\subfigure[$w=-1/3$ (F2)]{\includegraphics[width=0.2\textwidth]{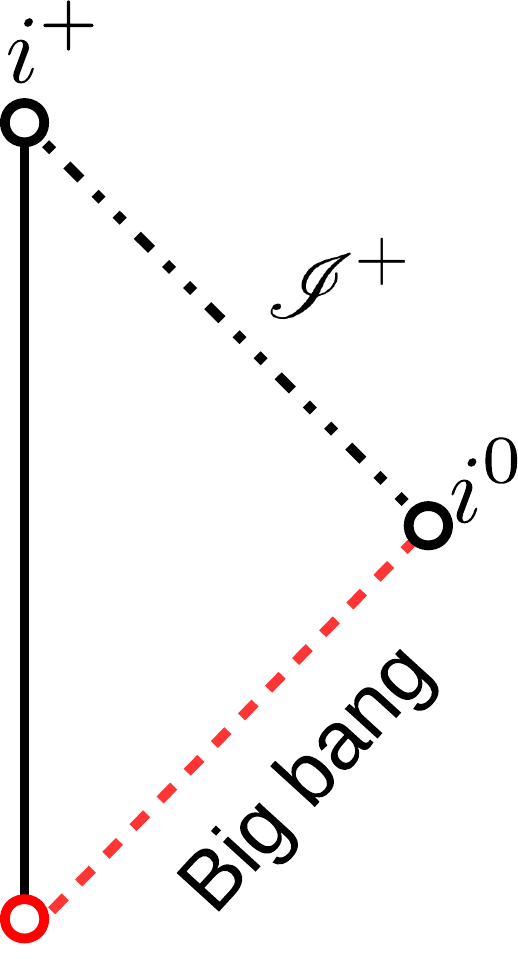}}
 & 
\subfigure[$-1<w<-1/3$ (F3)]{\includegraphics[width=0.2\textwidth]{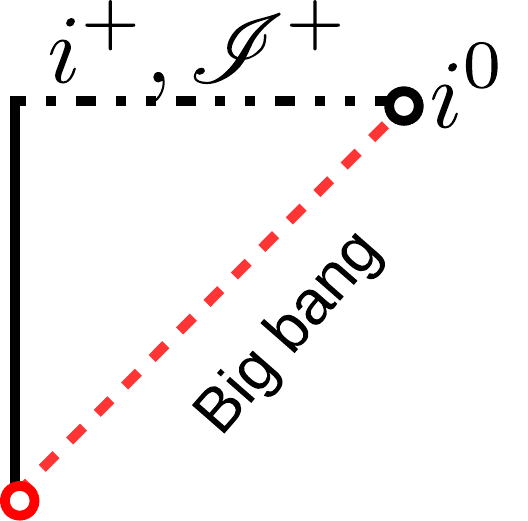}}
&
\subfigure[$w=-1$ (dS)]{\includegraphics[width=0.2\textwidth]{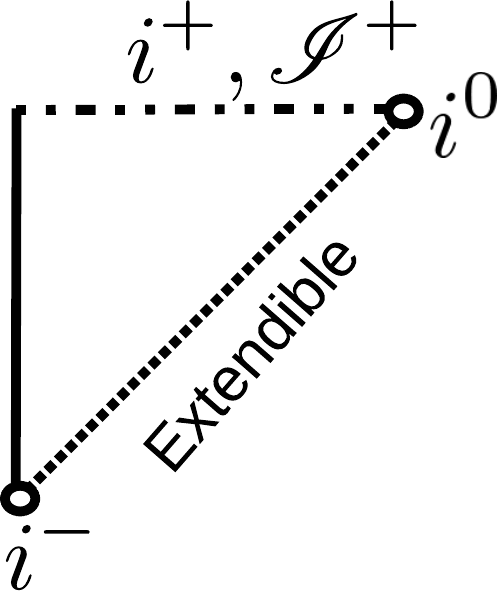}} \\
\subfigure[$-5/3<w<-1$ (F4a)]{\includegraphics[width=0.2\textwidth]{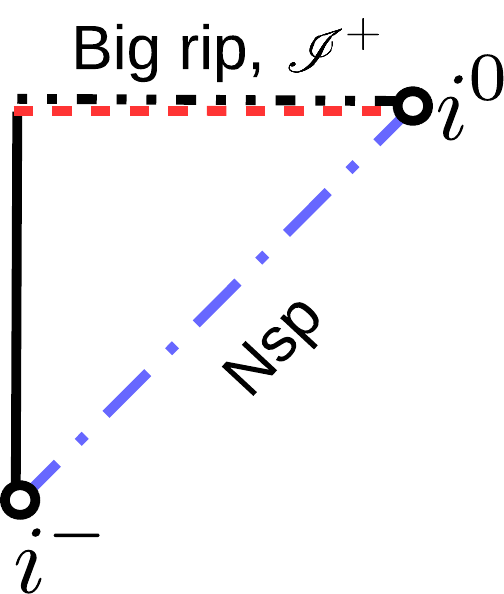}} &
\subfigure[$w=-5/3$ (F4b)]{\includegraphics[width=0.2\textwidth]{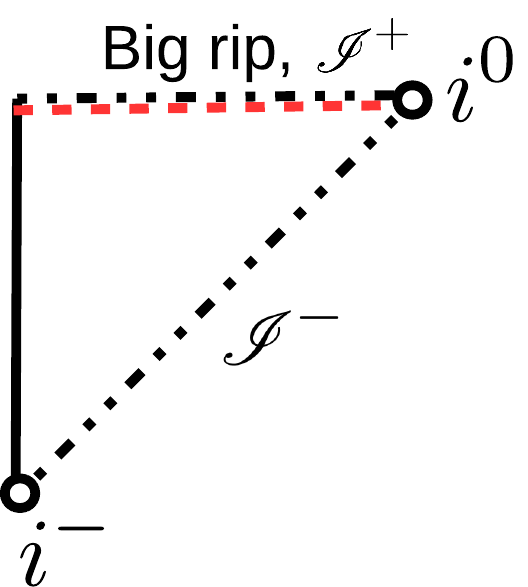}} &
\subfigure[$w<-5/3$ (F4c)]{\includegraphics[width=0.2\textwidth]{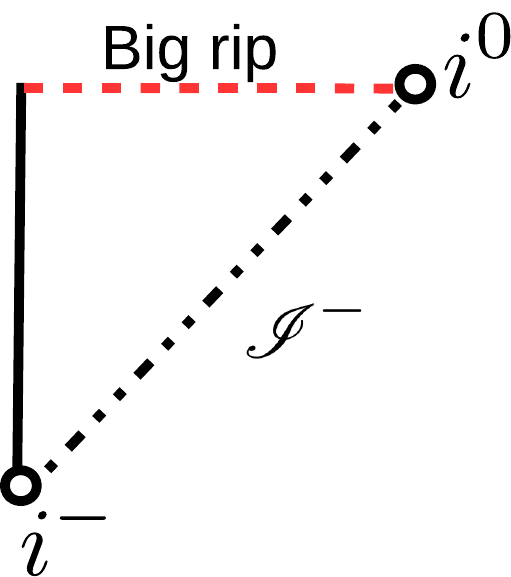}} & \quad 
 \end{tabular}
\caption{The Penrose diagrams for flat FLRW solutions with the positive density. \label{fg:flat_penrose}}
\end{center}
\end{figure}

\begin{figure}[H]
\begin{center}
 \begin{tabular}{cccc}
\subfigure[$w>-1/3$~(P1)]{\includegraphics[width=0.2\textwidth]{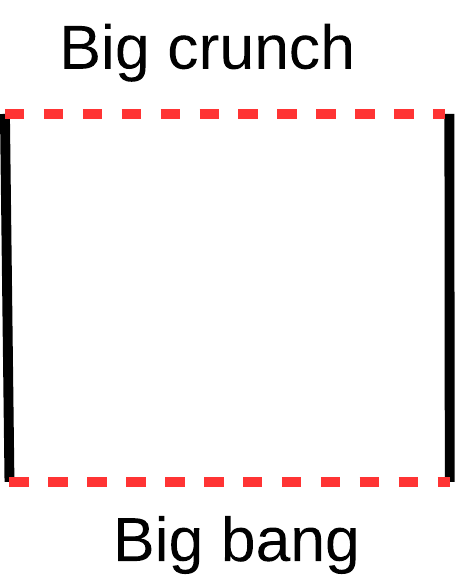}}
  & 
\subfigure[$w=-1/3$~(P2a)]{\includegraphics[width=0.2\textwidth]{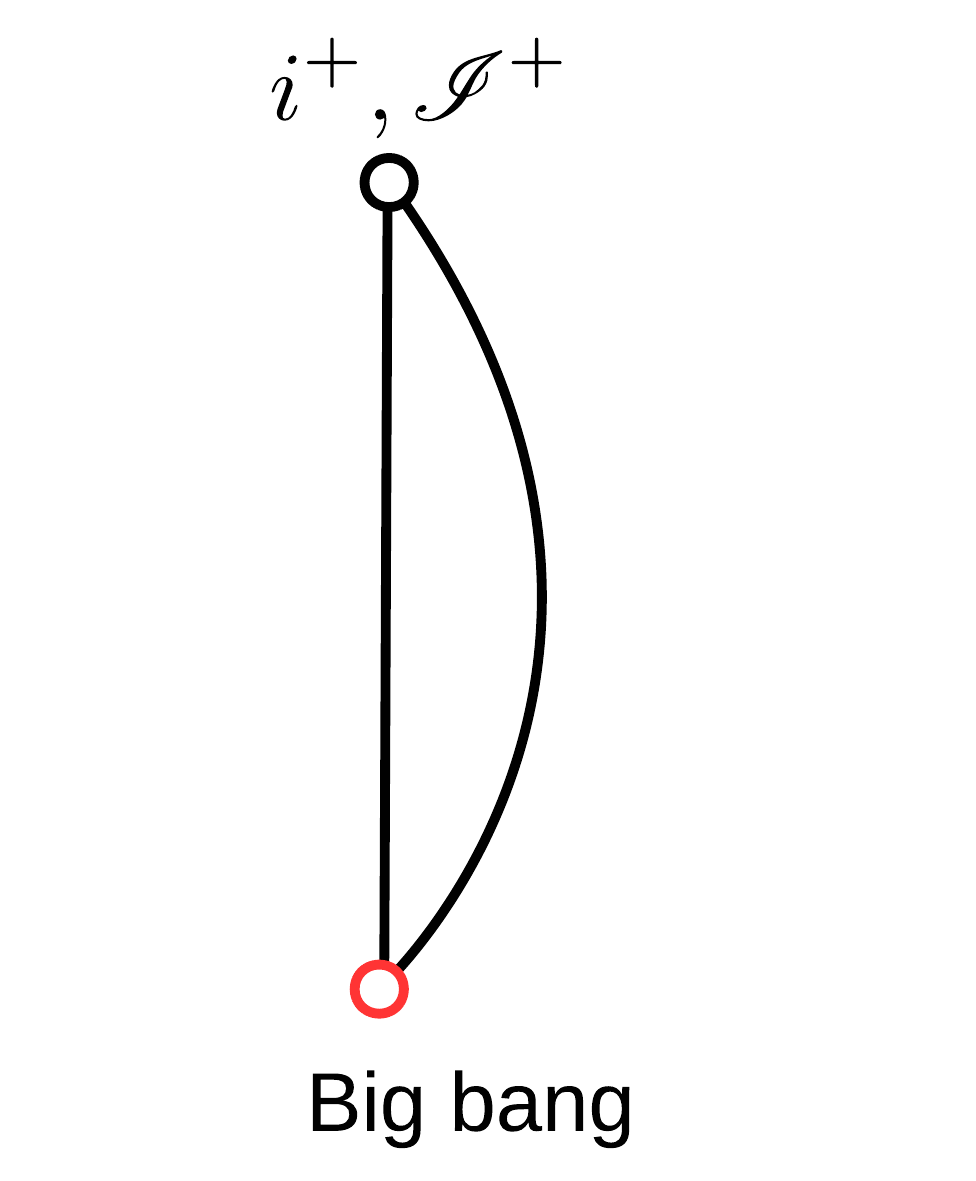}}
  &
\subfigure[$w=-1/3$~(P2b)]{\includegraphics[width=0.2\textwidth]{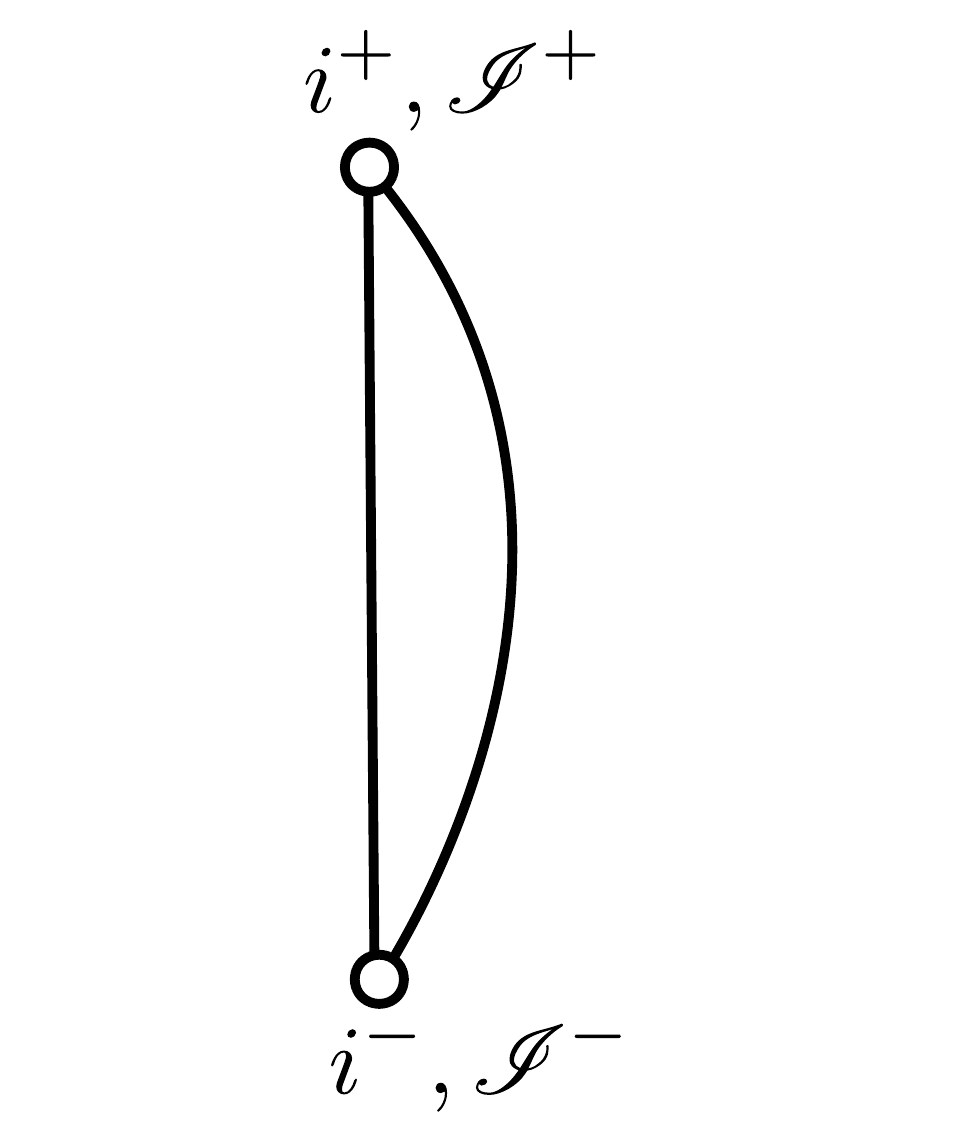}}
  &
\subfigure[$-1<w<-1/3$~(P3)]{\includegraphics[width=0.2\textwidth]{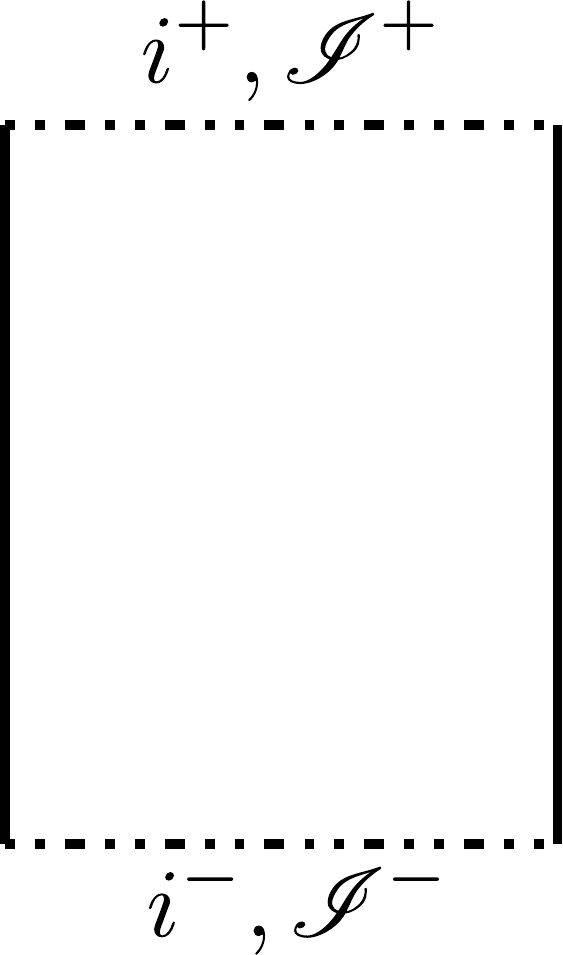}} 
\\
\subfigure[$w=-1$ (dS)]{\includegraphics[width=0.2\textwidth]{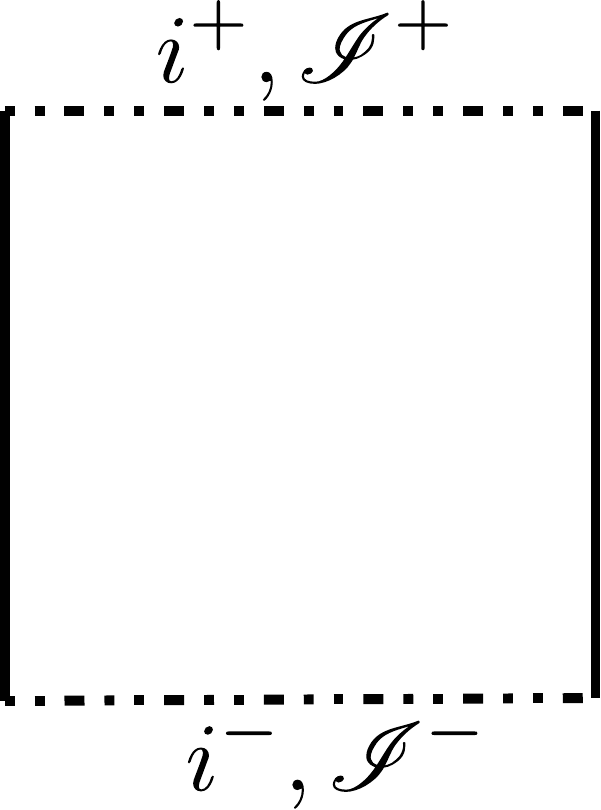}} &
\subfigure[$-5/3\le w<-1$ (P4a)]{\includegraphics[width=0.2\textwidth]{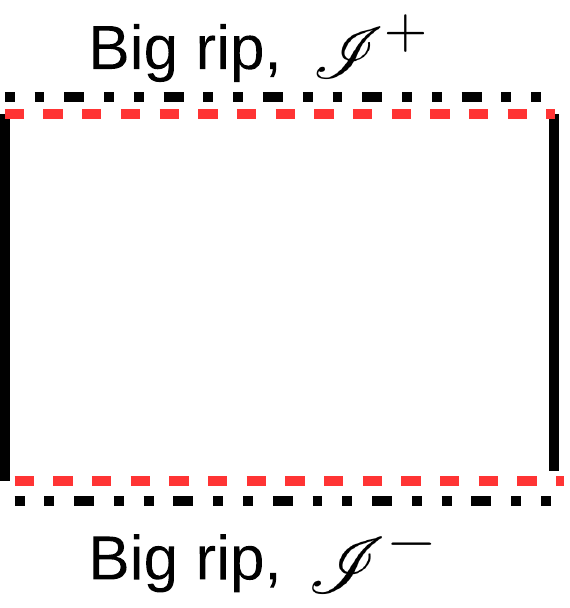}}
& 
\subfigure[$w<-5/3$ (P4b)]{\includegraphics[width=0.2\textwidth]{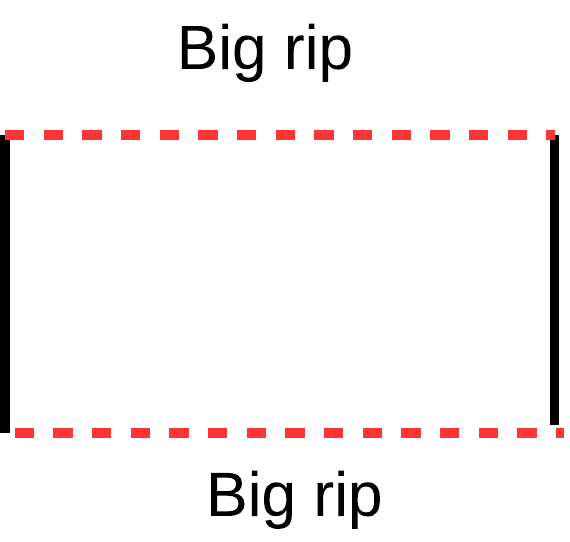}}&\quad 
 \end{tabular}
\caption{The Penrose diagrams for positive-curvature FLRW solutions. \label{fg:closed_penrose}}
\end{center}
\end{figure}

\begin{figure}[H]
\begin{center}
\includegraphics[width=0.2\textwidth]{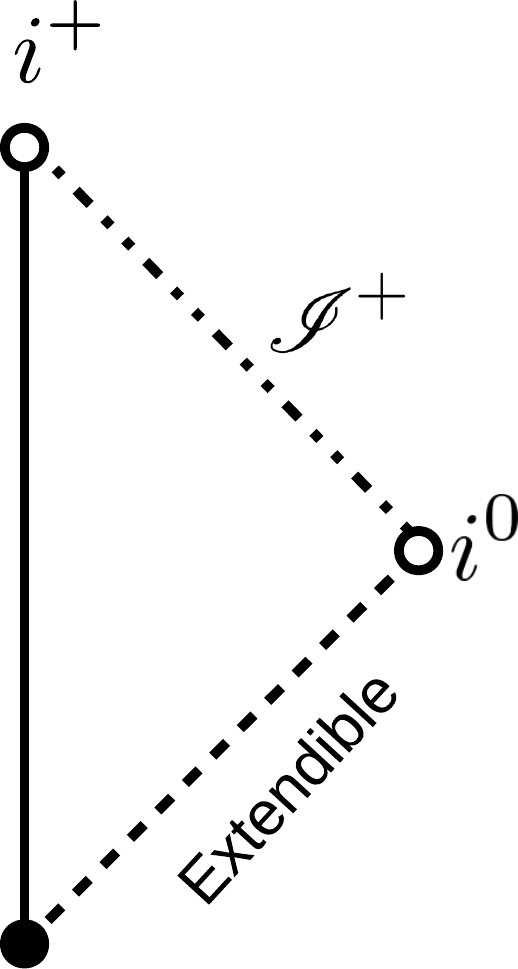}
\end{center}
\caption{The Penrose diagram for Milne spacetime. \label{fg:Milne_penrose}}
\end{figure}

\begin{figure}[H]
\begin{center}
 \begin{tabular}{cccc}
\subfigure[$w>-1/3$ (NP1)]{\includegraphics[width=0.2\textwidth]{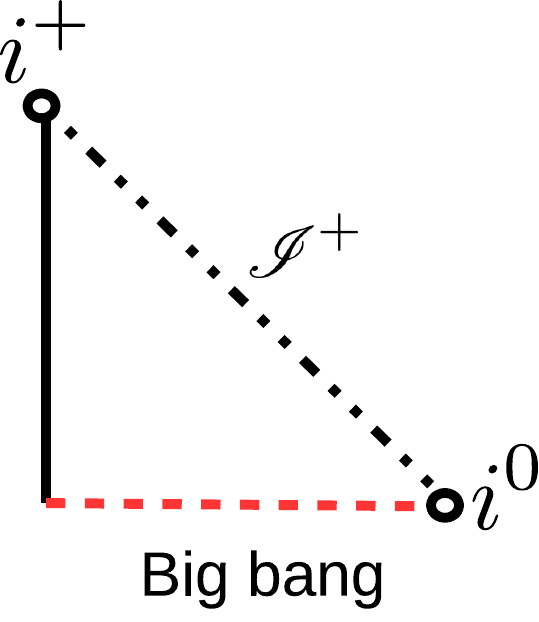}}
  & 
\subfigure[$w=-1/3$ (NP2)]{\includegraphics[width=0.2\textwidth]{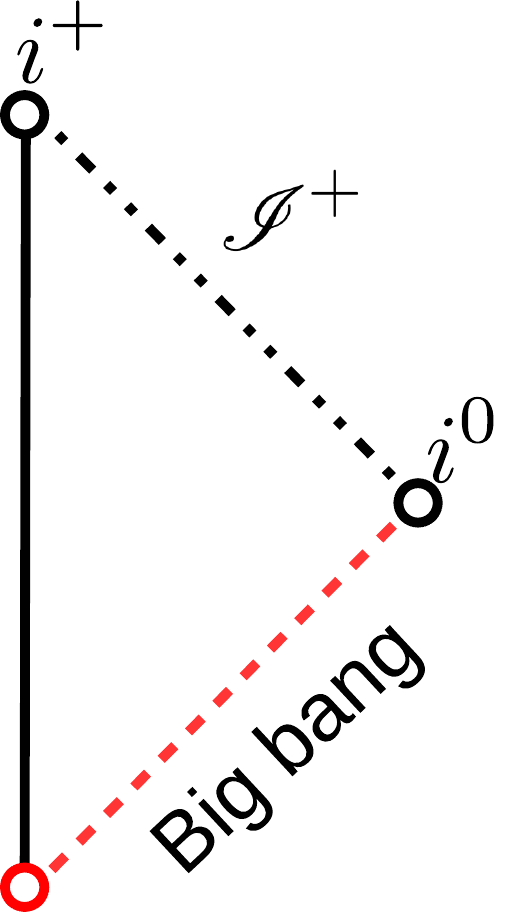}}
  &
\subfigure[$-1<w<-1/3$ (NP3)]{\includegraphics[width=0.2\textwidth]{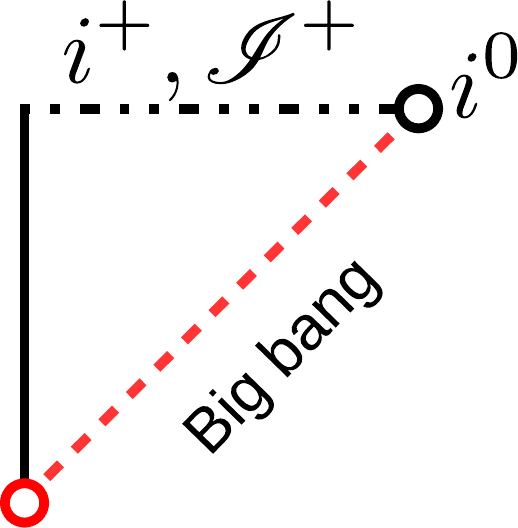}}
  &
\subfigure[$w=-1$ (dS)]{\includegraphics[width=0.2\textwidth]{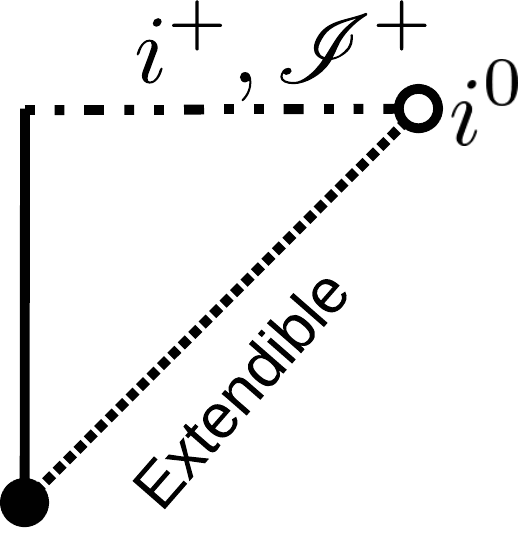}}
\\
\subfigure[$-5/3<w<-1$ (NP4a)]{\includegraphics[width=0.2\textwidth]{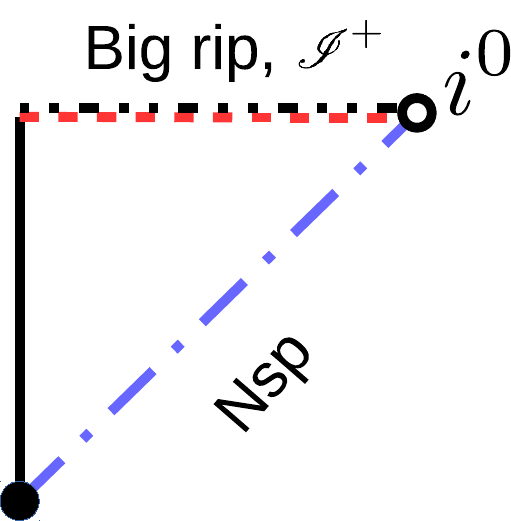}} 
&
\subfigure[$w=-5/3$ (NP4b)]{\includegraphics[width=0.2\textwidth]{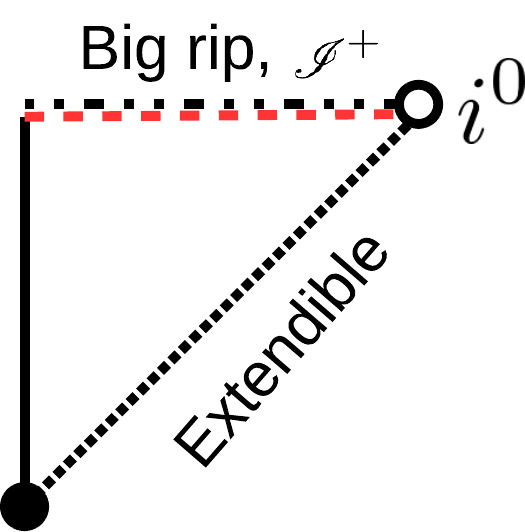}} 
&
\subfigure[$w<-5/3$ (NP4c)]{\includegraphics[width=0.2\textwidth]{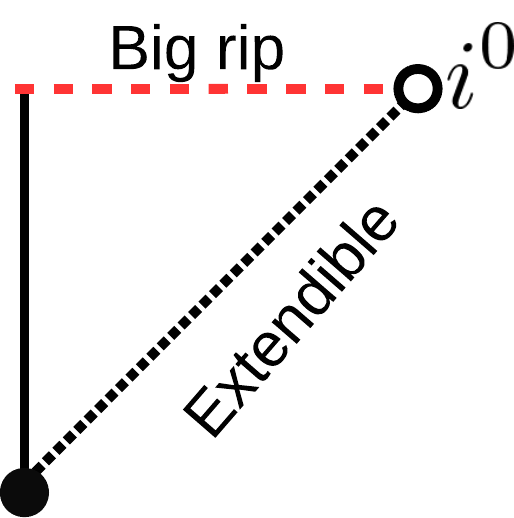}} 
&
\quad 
 \end{tabular}
\caption{The Penrose diagrams for negative-curvature FLRW solutions with the positive density. \label{fg:open_positive_penrose}}
\end{center}
\end{figure}

\begin{figure}[H]
\begin{center}
 \begin{tabular}{cccc}
\subfigure[$w>-1/3$ (NN1)]{\includegraphics[width=0.2\textwidth]{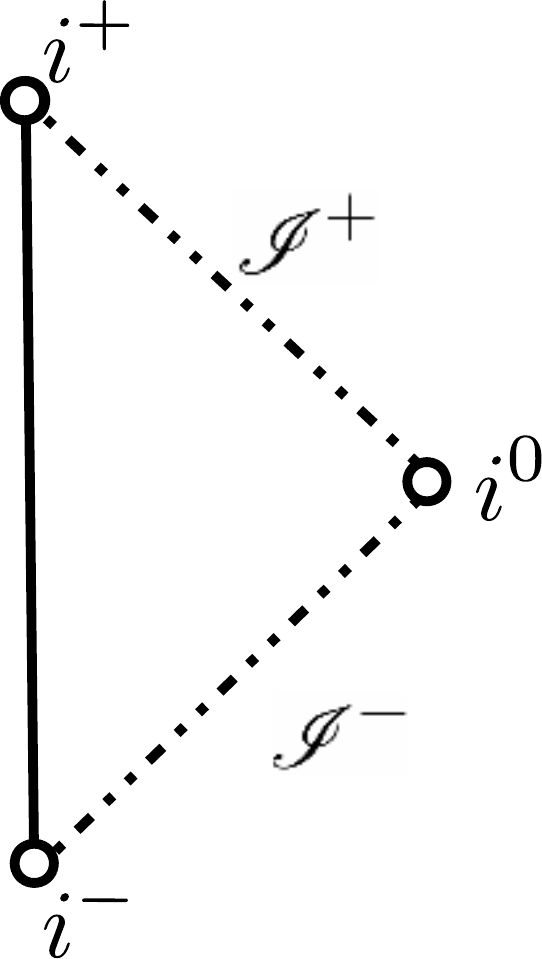}}
  & 
\subfigure[$w=-1/3$ (NN2a)]{\includegraphics[width=0.2\textwidth]{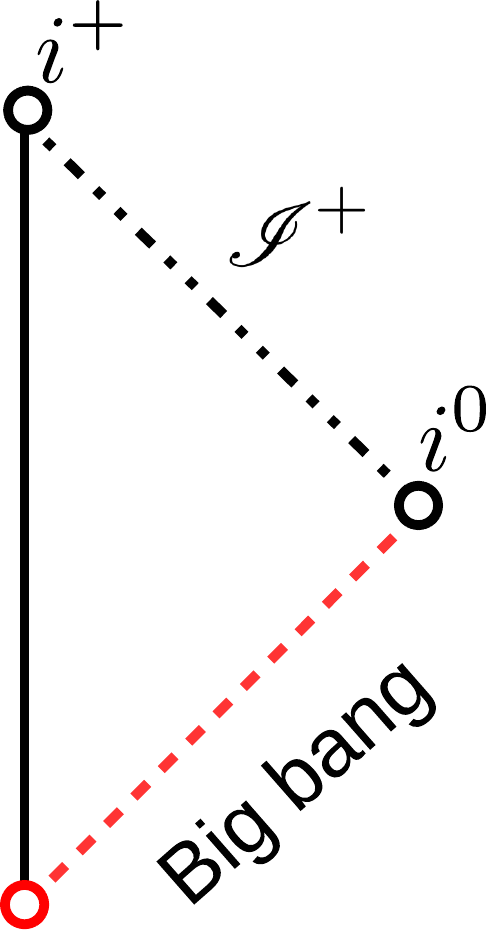}}
  &
\subfigure[$w=-1/3$ (NN2b)]{\includegraphics[width=0.2\textwidth]{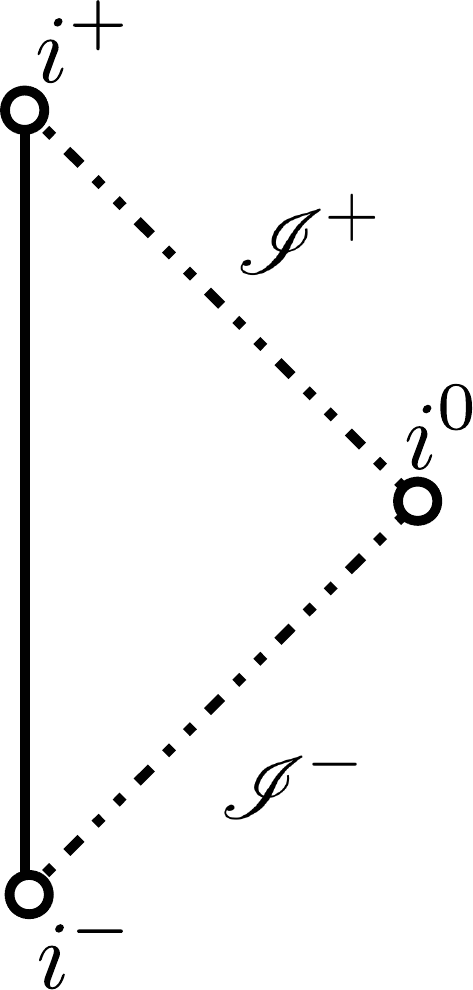}}
&
\subfigure[$-1<w<-1/3$ (NN3)]{\includegraphics[width=0.2\textwidth]{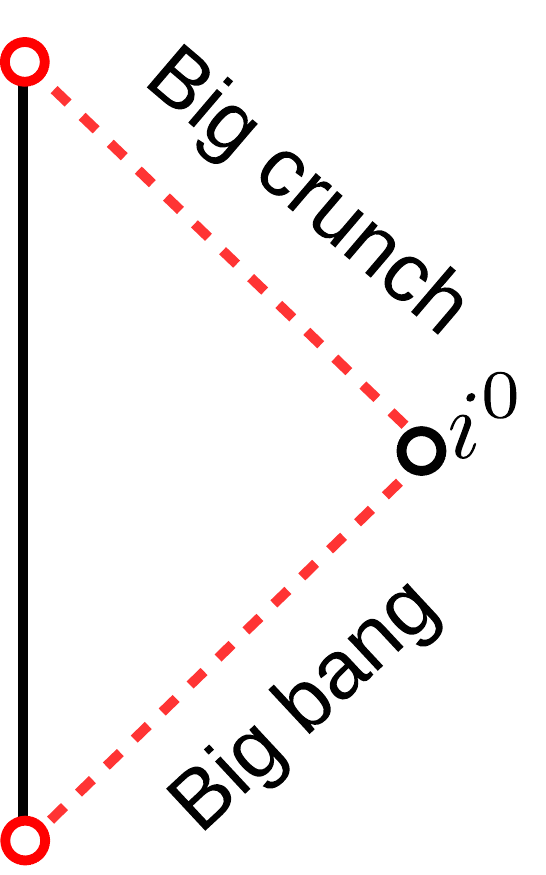}} \\
\subfigure[$w=-1$ (AdS)]{\includegraphics[width=0.2\textwidth]{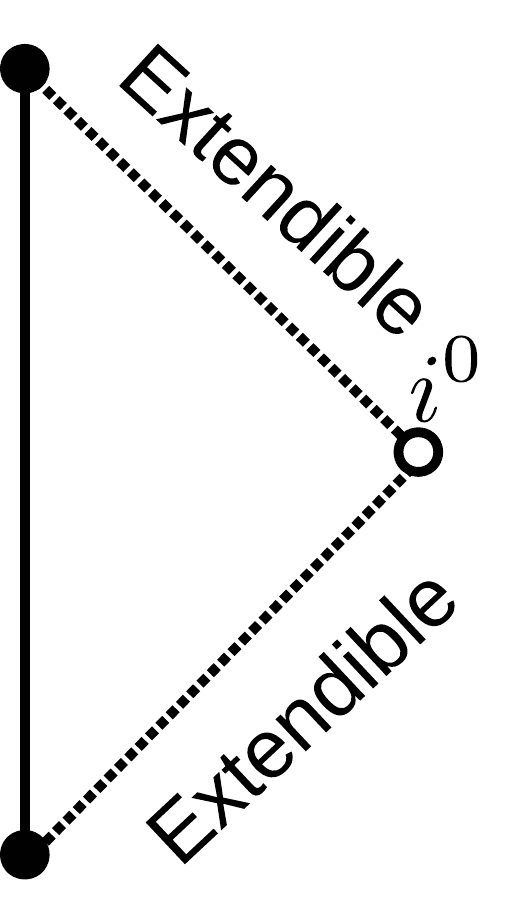}} 
&
\subfigure[$-5/3<w<-1$ (NN4a)]{\includegraphics[width=0.2\textwidth]{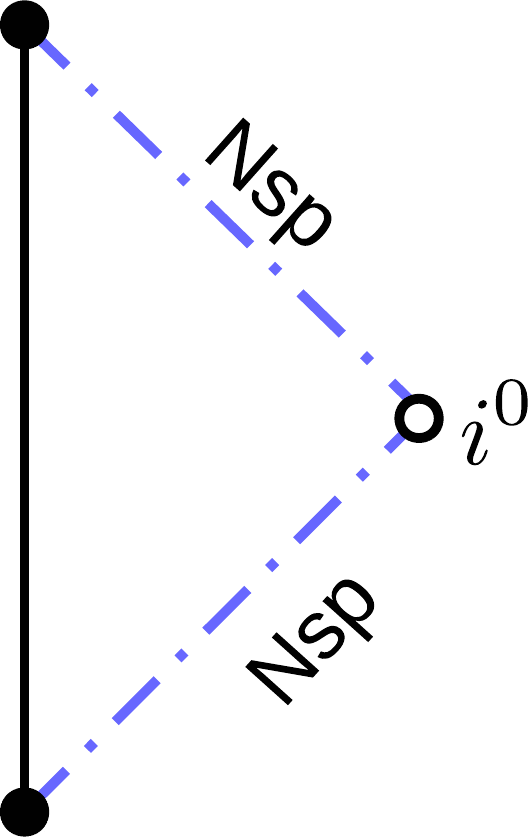}} 
&
\subfigure[$w\le -5/3$ (NN4b)]{\includegraphics[width=0.2\textwidth]{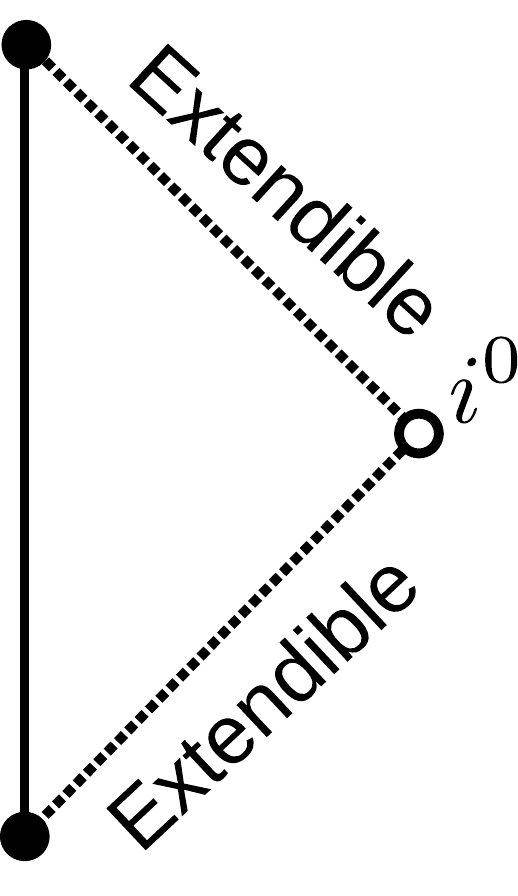}}
& 
\quad
 \end{tabular}
\caption{The Penrose diagrams for negative-curvature FLRW solutions with the negative density. \label{fg:open_negative_penrose}}
\end{center}
\end{figure}

Before proceeding to technicalities, we 
summarise our main results.
The conformal structure of the FLRW solutions depends on the signs of 
$K$ 
and 
$\rho$; it also depends on
$w$,  together with an integration constant in the special case 
$w=-1/3$.
The Penrose diagrams for the flat case with $\rho=0$ (Minkowski spacetime) and $\rho>0$ are shown in figures~\ref{fg:Minkowski_penrose} and \ref{fg:flat_penrose}, respectively.
The black solid, black dotted, black dashed-double-dotted, 
blue dashed-dotted 
and red dashed lines denote regular centres, extendible conformal boundaries, 
line-like conformal infinities, non-s.p. curvature singularities and 
s.p. curvature singularities, respectively, and the 
black filled, black open and red open circles denote point-like 
extendible boundaries, infinities and s.p. curvature singularities, 
respectively.
The past null boundary for the flat case 
is a non-s.p. curvature singularity for $-5/3<w<-1$.
The Penrose diagrams for the positive-curvature case are shown in figure~\ref{fg:closed_penrose}.
Those
 for the negative-curvature case with $\rho=0$ (Milne spacetime), $\rho>0$ and $\rho<0$ are shown in figures~\ref{fg:Milne_penrose}, \ref{fg:open_positive_penrose} and \ref{fg:open_negative_penrose}, respectively.
For the negative-curvature case 
with $-5/3<w<-1$, 
the past null boundary with $\rho>0$ is a non-s.p. curvature singularity;
this also applies for
the future and past null boundaries with $\rho<0$.


The Penrose diagrams for
the flat FLRW solutions with
$-1<w<1$,   
investigated in
reference~\cite{Senovilla:1998oua},  have a big-bang spacelike singularity 
for $-1/3<w<1$ and a big-bang null singularity for $-1<w< -1/3$ 
at the past boundary; they also have a future null infinity at the future boundary.
For $w\le -1/3$, 
there is unexpectedly rich structure
with a variety of past and future 
boundaries.
For $\rho>0$ and $w<-1$, the spacetime commonly ends with a future big-rip spacelike singularity, although 
for $-5/3 \le w<-1$,
it is still future null geodesically complete. 
This 
 is consistent with reference~\cite{Fernandez-Jambrina:2006tkb}.
In the flat and negative-curvature cases with $\rho>0$,
the expanding Universe begins with a big-bang spacelike singularity for $w>-1/3$, a big-bang null singularity for $-1<w\le -1/3$
and a non-s.p. curvature null singularity for $-5/3<w<-1$.
This implies that no quasi-exponential inflationary model
avoids the initial singularity problem if it is described by 
the flat 
or negative-curvature solutions with $-5/3<w<-1$ or $-1<w<-1/3$.
For $w\le -5/3$, the flat and negative-curvature solutions have a past null infinity and 
a past extendible null boundary, respectively.
For $K=1$ and $w<-1/3$, the solution describes a bouncing Universe, which is singularity-free for $-1\le w<-1/3$
but has past and future big-rip singularities for $w<-1$.
For $w>-5/3$ and $\rho>0$, no spacelike geodesic emanates from or terminates at 
the spacelike infinity in the flat and negative-curvature cases.
Although the nature of the matter 
corresponding to $w=-5/3$ is unclear, 
the Einstein equation suggests that this value is just as critical as 
$1/3$, $-1/3$ and $-1$.

The paper is organised as follows. 
Section II gives the conformal completion of the FLRW spacetimes and the 
corresponding solutions of the Einstein equation 
for $p=w\rho $. Section III 
derives
expressions for general geodesics and 
the Ricci tensor components in the p.p.
frame.  Section IV 
provides the definition of singularities and infinities as conformal boundaries. We analyse the conformal boundaries of the FLRW solutions 
and infer the Penrose diagrams. 
Appendix~\ref{sec:abbreviations} gives a
list of abbreviations and Appendix~\ref{sec:integrals}
gives some important integrals.
Henceforth we use 
units
with $c=G=1$ 
and the abstract index notation in Wald's textbook~\cite{Wald:1984rg}.

\section{The FLRW spacetime}
\label{sec:FLRW_spacetime}

\subsection{Conformal completion}

The line element in the FLRW spacetime is written in the following form:
\begin{eqnarray}
 ds^{2}&=&-dt^{2}+a^{2}(t)[dr^{2}+\Sigma_{K}^{2}(r)d\Omega^{2}] \nonumber \\
&=&a^{2}(\eta)[-d\eta^{2}+dr^{2}+\Sigma_{K}^{2}(r)d\Omega^{2}],
\end{eqnarray}
where $a>0$, $ad\eta =dt$, $d\Omega^{2}=d\theta^{2}+\sin^{2}\theta d\phi^{2}$ with $0\le \theta \le \pi$ and $0\le \phi <2\pi$, and 
\begin{eqnarray}
 \Sigma_{K}(r)=\left\{\begin{array}{cc}
  r & (K=0)\\
  \sin r & (K=1) \\
  \sinh r & (K=-1)
		  \end{array}
\right. .
\label{eq:areal_radius}
\end{eqnarray}
The domain of the coordinates $(r,\eta)$ depends on the spatial curvature and is given by
\begin{equation}
 -\infty<\eta<\infty,~ \left\{\begin{array}{cc}
  0\le r<\infty ~
& (K=0,-1)\\
  0\le r\le \pi  & (K=1) \\
		  \end{array}
\right. 
\end{equation}
unless $\eta$ is 
restricted to a smaller interval by the specific solution.

\begin{figure}[H]
\begin{center}
\begin{tabular}{ccc}
 \subfigure[\label{fg:flat_domain}]{\includegraphics[width=0.25\textwidth]{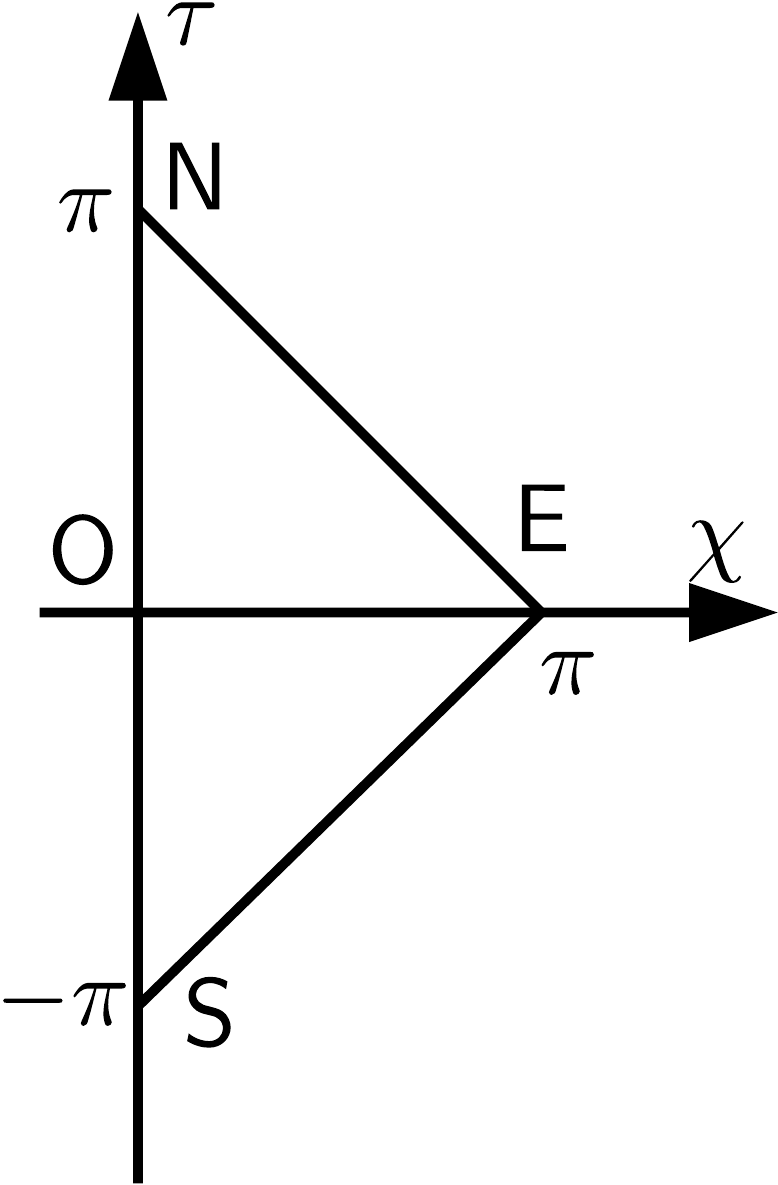}}
& 
 \subfigure[\label{fg:closed_domain}]{\includegraphics[width=0.25\textwidth]{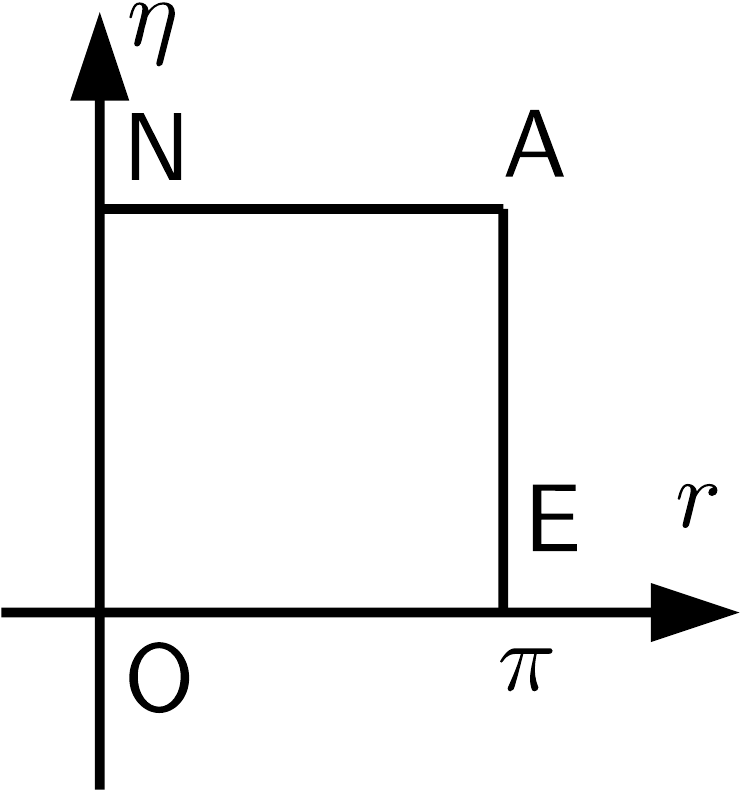}}
&
 \subfigure[\label{fg:closed_exception_domain}]{\includegraphics[width=0.25\textwidth]{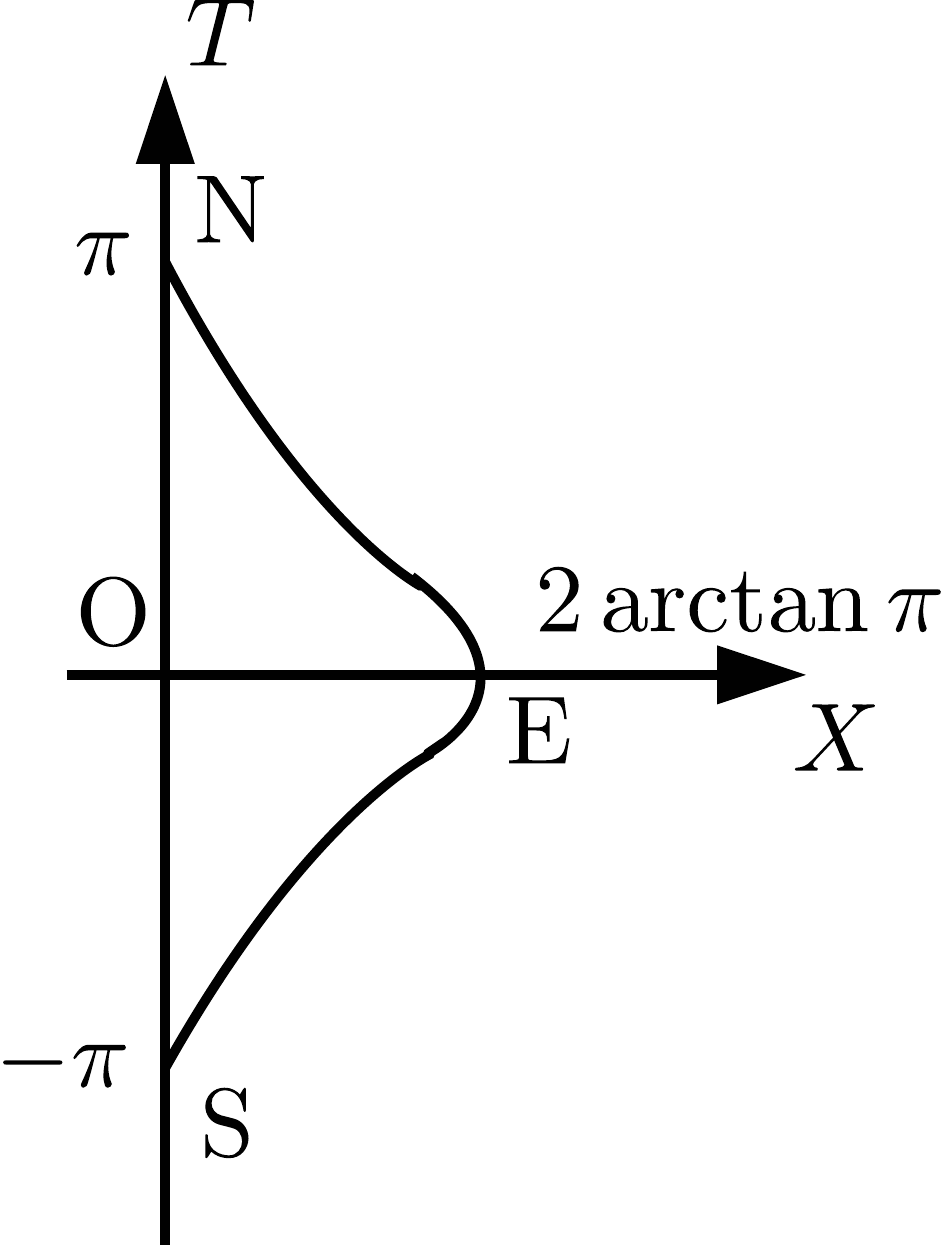}}
\\
 \subfigure[\label{fg:closed_domain_2}]{\includegraphics[width=0.25\textwidth]{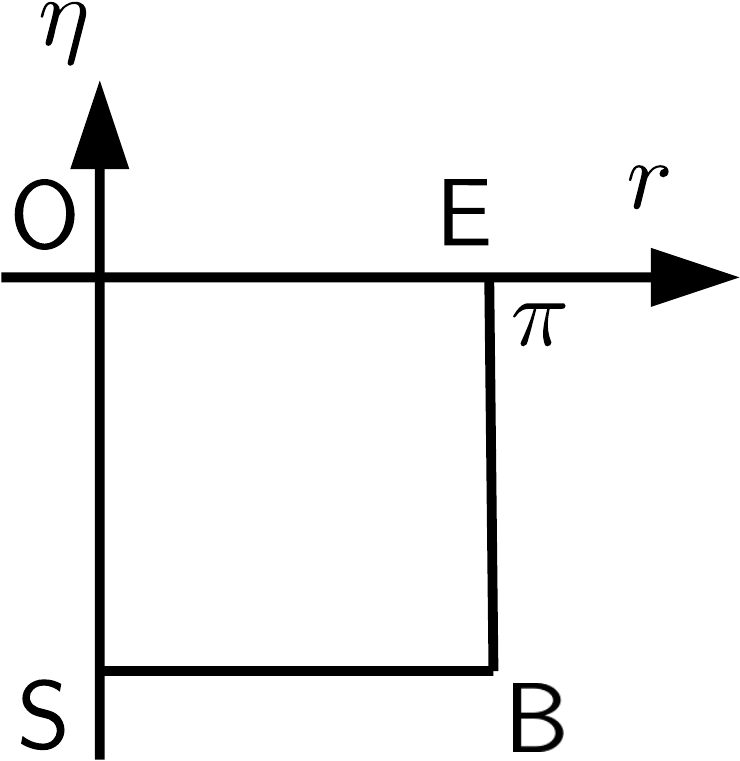}}
& 
 \subfigure[\label{fg:open_domain}]{\includegraphics[width=0.25\textwidth]{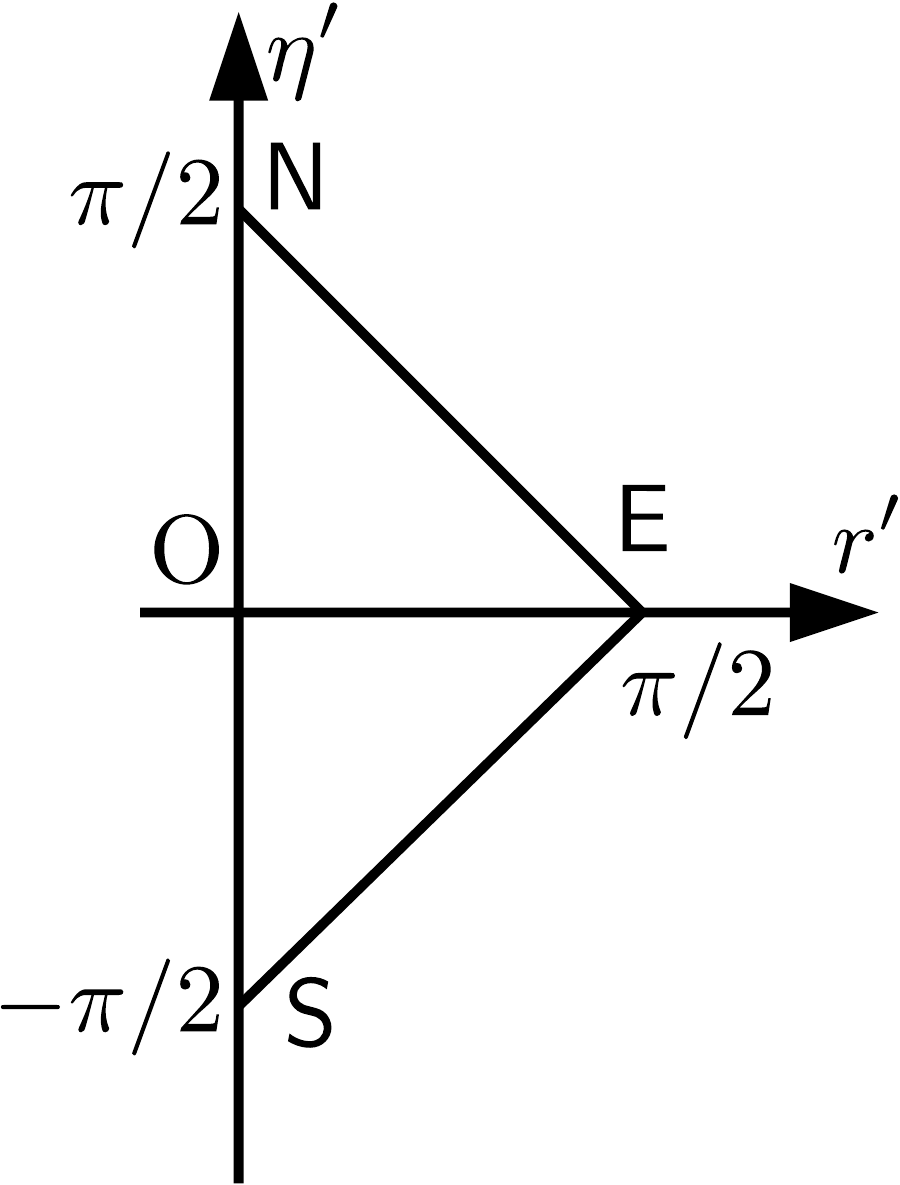}}
&
\end{tabular}
\caption{The domain of the coordinate planes for (a) the flat case, the positive-curvature cases for (b) $w>-1/3$, (c) $w=-1/3$ and (d) $w<-1/3$, and (e) the negative-curvature case.}
\end{center} 
\end{figure}

The conformal completion 
prescribes the ``boundary'' of the spacetime $M$.   If this
is conformally isometric to a bounded open region of the ``unphysical'' spacetime $\tilde{M}$, the boundary of the image of $M$ in $\tilde{M}$ is called the conformal boundary~\cite{Hawking:1973uf,Wald:1984rg}.
The line element in the flat FLRW spacetime can be rewritten as
\begin{equation}
ds^{2}=\frac{1}{4}a^{2}(\eta)\sec^{2} \left( \frac{\tau+\chi}{2} \right) \sec^{2} \left(\frac{\tau-\chi}{2} \right)
(-d\tau^{2}+d\chi^{2}+\sin^{2}\chi d\Omega^{2}), 
\end{equation}
where 
\begin{equation}
 \tau=\arctan(\eta+r)+\arctan(\eta-r),\quad \chi=\arctan(\eta+r)-\arctan(\eta-r). 
\label{eq:tauchi_etar}
\end{equation}
Thus, the physical spacetime is conformally isometric to the bounded region 
$ \chi \ge 0$, $\tau-\chi>- \pi$ and $\tau+\chi< \pi$
in the Einstein static Universe.
The domain of $(\chi,\tau)$ is 
the isosceles right-angled triangle
in figure~\ref{fg:flat_domain}. We denote the top and bottom vertices, the apex and the 
midpoint of the base by
{\sf N}, {\sf S}, {\sf E} and {\sf O}, respectively. The coordinates $(\chi,\tau)$ of {\sf N}, {\sf S}, {\sf E} and {\sf O} 
are $(0,\pi)$, $(0,-\pi)$, $(\pi,0)$ and $(0,0)$, respectively.

For the positive-curvature case, the line element is already 
conformal to that of the Einstein static Universe.  As we will see in Sec.~\ref{sec:FLRW_solutions}, 
the domain of $\eta$ 
depends on the parameter $w$. 
For $w>-1/3$ and $w<-1/3$, the domains 
are restricted to the bounded intervals $0<\eta<2\pi/(1+3w)$ and $2\pi/(1+3w)<\eta<0$,
respectively. Since the domain of $r$ is also bounded, the domains of $(r,\eta)$ 
for $w>-1/3$ and $w<-1/3$
are both given by rectangles, as depicted
in figures~\ref{fg:closed_domain} and \ref{fg:closed_domain_2}, respectively,
so no conformal rescaling is needed. 
For $w>-1/3$, 
we denote the top left, top right, bottom left and bottom right vertices by {\sf N}, {\sf A}, {\sf O} and {\sf E}, respectively,  and their coordinates $(r,\eta)$ 
are $(0,2\pi/(1+3w))$, $(\pi,2\pi/(1+3w))$, $(0,0)$ and $(\pi,0)$, respectively.
For $w<-1/3$, we denote the top left, top right, bottom left and bottom right vertices by
 {\sf O}, {\sf E}, {\sf S} and {\sf B}, respectively, 
with coordinates
$(0,0)$, $(\pi,0)$, $(0,2\pi/(1+3w))$ and $(\pi,2\pi/(1+3w))$, respectively.
However,  in the exceptional case $w=-1/3$,  the domain of $\eta$ is unbounded
($-\infty<\eta<\infty$) and we can use the line element 
\begin{eqnarray}
  ds^{2}&=&\frac{1}{4}a^{2}(\eta)\sec^{2}\frac{T+X}{2} \sec^{2}\frac{T-X}{2}\nonumber \\
&\times& \left(-dT^{2}+dX^{2}+4\cos^{2}\frac{T+X}{2}\cos^{2}\frac{T-X}{2}\sin^{2}r d\Omega^{2}\right),
\end{eqnarray}
where 
\begin{equation}
T=\arctan (\eta+r)+\arctan(\eta-r),\quad X=\arctan (\eta+r)-\arctan(\eta-r).
\end{equation}
Thus, the physical spacetime is conformally isometric to the bounded region 
\begin{equation}
 X\ge 0, -\pi <T<\pi,~ \tan\frac{T+X}{2}-\tan\frac{T-X}{2}\le 2\pi \, ,
\end{equation}
as depicted 
in figure~\ref{fg:closed_exception_domain}. 
Here 
we denote the top, bottom and right vertices by {\sf N}, {\sf S} and {\sf E}, respectively, 
and the midpoint of the line segment $\overline{\sf NS}$ by {\sf O}. The  
coordinates $(X,T)$ of {\sf N}, {\sf S}, {\sf E} and {\sf O} are 
$(0,\pi)$, $(0,-\pi)$, $(2\arctan\pi,0)$ and $(0,0)$, respectively.

For the negative-curvature case, the line element can be rewritten as
\begin{equation}
 ds^{2}=\frac{a^{2}(\eta)}{\cos(r'+\eta')\cos(r'-\eta')}(-d\eta'^{2}+dr'^{2}+\sin^{2}r' d\Omega^{2}),
\end{equation}
where
\begin{equation}
 \tan \eta'=\frac{\sinh\eta}{\cosh r},\quad \tan r'=\frac{\sinh r}{\cosh \eta}~.
\end{equation}
The original domain is mapped to the isosceles right-angled triangle 
\begin{equation}
r'\ge 0,\quad  \eta'-r'>-\frac{\pi}{2},\quad \eta'+r'<\frac{\pi}{2}.
\end{equation}
Thus, the physical spacetime is conformally isometric to a bounded 
region in the Einstein static Universe as depicted in figure~\ref{fg:open_domain}, 
where we donote the top and bottom vertices, the apex and the midpoint of the base by {\sf N}, {\sf S}, {\sf E} and {\sf O}, respectively, and 
the coordinates $(r',\eta')$ of {\sf N}, {\sf S}, {\sf E} and {\sf O} are 
$(0,\pi/2)$, $(0,-\pi/2)$, $(\pi/2,0)$ and $(0,0)$, respectively.

\subsection{FLRW solutions with $p=w\rho$}
\label{sec:FLRW_solutions}

The Einstein equation implies that the matter field 
must have the 
perfect-fluid form
\begin{equation}
 T^{ab}=(\rho+p) u^{a}u^{b}+p g^{ab},
\end{equation}
where $\rho$, $p$ and 
$u^{a}:=a^{-1}(\partial/\partial \eta)^{a}$ are the 
energy density, pressure and four-velocity of the fluid element, respectively.
If we further assume $p=w\rho$ with $w=\mathrm{const}$, the conservation law $\nabla^{a}T_{ab}=0$ implies 
\begin{equation}
\rho=\rho_{0}\left(\frac{a_{0}}{a}\right)^{3(1+w)},
\end{equation}
where $\rho=\rho_{0}$ when $a=a_{0}$.
For $w\ne -1/3$, the Einstein equation reduces to 
\begin{equation}
 \left(\frac{d\tilde{a}}{d\tilde{t}}\right)^{2}=\frac{\tilde{a}_{c}}{\tilde{a}}-K,
\label{eq:Hubble}
\end{equation}
where
\begin{equation}
 \tilde{a}:=a^{1+3w}, ~~d\tilde{t}:=(1+3w)\tilde{a}^{3w/(1+3w)}dt, ~~\tilde{a}_{c}:=\frac{8\pi }{3}\rho_{0}a_{0}^{3(1+w)}. 
\end{equation}
 For $w=-1/3$, it 
becomes
\begin{equation}
\left(\frac{da}{dt}\right)^{2}=\tilde{a}_{c}-K,
\label{eq:Hubble_minus_third}
\end{equation}
where
\begin{equation}
 \tilde{a}_{c}:=\frac{8\pi }{3}\rho_{0}a_{0}^{2}. 
\end{equation}

For the vacuum case, only $K=0$ and $K=-1$ are possible. 
For $K=0$, the solution is $a=a_{0}=\mathrm{const}$, 
corresponding to Minkowski spacetime, while for $K=-1$, it is $a=b_{0}e^{\eta}$
with $b_{0}$ being a positive constant, corresponding to Milne spacetime. The domain of $\eta$ is $-\infty<\eta<\infty$ for both of the cases.

For the non-vacuum case, we first consider $w\ne -1/3$.
For $K=0$ and $\rho>0$, equation~(\ref{eq:Hubble}) can be integrated to give
\begin{equation}
 a(\eta)= 
\begin{cases}
b_{0} \eta^{\alpha}& \quad \mbox{for}~~0<\eta <\infty ~~ (w>-1/3) \\
b_{0} (-\eta)^{\alpha} & \quad \mbox{for}~~-\infty<\eta<0 ~~ (w<-1/3) 
\end{cases}
,
\label{eq:scale_factor_flat_FLRW}
\end{equation}
where $\alpha := 2/(1+3w)$, $b_{0}$ is a positive constant and only an expanding branch is focused.
This class contains the Einstein-de Sitter Universe ($w=0$) and de Sitter spacetime 
in the flat chart ($w=-1$). 
For $K=1$, the energy density must be positive and equation~(\ref{eq:Hubble}) can be integrated to give
\begin{eqnarray}
 \tilde{a}=\tilde{a}_{c} \frac{1-\cos\tilde{\eta}}{2} \, ,\quad \tilde{t}=\tilde{a}_{c} \frac{\tilde{\eta}-\sin\tilde{\eta}}{2}
\end{eqnarray}
with $\eta=\tilde{\eta}/(1+3w)$, which contains de Sitter spacetime in the global chart ($w=-1$). The domain of $\eta$ is thus $0<\eta<2\pi/(1+3w)$ and $2\pi/(1+3w)<\eta<0$
for $w>-1/3$ and $w<-1/3$, respectively.
For $K=-1$ and $\rho>0$, equation~(\ref{eq:Hubble}) can be integrated to give
\begin{eqnarray}
 \tilde{a}=\tilde{a}_{c} \frac{\cosh\tilde{\eta}-1}{2} \, ,~\tilde{t}=\tilde{a}_{c} \frac{\sinh\tilde{\eta}-\tilde{\eta}}{2} \, ,
\end{eqnarray}
which contains de Sitter spacetime in the 
open chart ($w=-1$).
The domain of $\eta$ is $0<\eta<\infty$ and $-\infty<\eta<0$ for $w>-1/3$ and $w<-1/3$, respectively.
For $K=-1$ and $\rho<0$, 
\begin{eqnarray}
 \tilde{a}=\tilde{a}'_{c}
  \frac{1+\cosh\tilde{\eta}}{2},~~\tilde{t}=\tilde{a}'_{c}
  \frac{\tilde{\eta}+\sinh\tilde{\eta}}{2} \,  
\end{eqnarray}
with $\tilde{a}_{c}'=-\tilde{a}_{c}$, which contains anti-de Sitter spacetime 
in the open chart ($w=-1$).
The domain of $\eta$ is $-\infty<\eta<\infty$.

For the $w=-1/3$ non-vacuum case, 
equation~(\ref{eq:Hubble_minus_third})
again implies that the energy density can be negative only for $K=-1$.
For $\tilde{a}_{c}-K>0$, equation~(\ref{eq:Hubble_minus_third}) can be integrated to give
$
a=b_{0}e^{b_{c}\eta}, 
$
with $b_{0}$ being a positive constant and $b_{c}=\sqrt{\tilde{a}_{c}-K}$, 
while for $\tilde{a}_{c}-K=0$, the solution is static 
and
$
 a=a_{0}=\mbox{const}.
$
The domain of $\eta$ is $-\infty<\eta<\infty$ for both of the cases.

\section{Geodesics, spacelike curves and the Riemann tensor}

\subsection{Geodesics and spacelike curves}
Due to the symmetry, without loss of generality we 
can consider geodesics in the two-dimensional timelike plane
$\theta=\pi/2$ and $\phi=0$.
 The Lagrangian then takes the simplified form 
\begin{equation}
L=\frac{1}{2}a^{2}(\eta)(-\dot{\eta}^{2}+\dot{r}^{2}),
\end{equation}
where the dot denotes differentiation with respect to the affine parameter $\lambda$.
The Euler-Lagrange equations are 
\begin{eqnarray}
\frac{d}{d\lambda}(a^{2}\dot{\eta})+aa'(-\dot{\eta}^{2}+\dot{r}^{2})&=&0 \label{eq:EL0}, \\
\frac{d}{d\lambda}(a^{2}\dot{r})&=&0 \label{eq:EL1},
\end{eqnarray}
where the prime denotes 
differentiation
with respect to $\eta$.
The above equations 
can be  integrated to give
\begin{eqnarray}
\dot{r}&=&\frac{C}{a^{2}}, \label{eq:rdot}\\
a^{2}(-\dot{\eta}^{2}+\dot{r}^{2})&=&D \label{eq:normalisation} \, , 
\end{eqnarray}
where $C$ and $D$ are constants. We can rescale $\lambda$ so that 
$D=1$ and $-1$ for the spacelike and timelike geodesics, respectively, while $D=0$
for the null geodesics.
Clearly, $r$ has no turning point.

If we take $\eta=\eta(\lambda)$, equations~(\ref{eq:rdot}) and (\ref{eq:normalisation}) reduce to
the one-dimensional potential form
\begin{equation}
 \dot{\eta}^{2}+V_{\eta}(\eta)=0\quad {\rm with} \quad V_{\eta}(\eta)=-\frac{C^{2}}{a^{4}(\eta)}+\frac{D}{a^{2}(\eta)}.
\label{eq:Vpotential}
\end{equation}
This implies that $\eta$ has no turning point for $D=0,-1$, while
the allowed region is given by $a(\eta)\le |C|$ for $D=1$ 
with a turning point at $\eta=\eta_{\rm tp}$ where $a(\eta_{\rm tp})=|C|$.
As $r=0$ is a regular centre, $C$ becomes $-C$ as the geodesic passes through there.
We classify geodesics as follows.
\begin{itemize}
 \item[(i)] Null geodesic ($D=0$). If we take the affine parameter to have $\dot{\eta}>0$,
equation~(\ref{eq:normalisation}) 
can be integrated to give 
\begin{equation}
 \eta-\eta_{0}=\pm (r-r_{0}),
\end{equation}
where $\eta=\eta_{0}$ when $r=r_{0}$.
We can make $C=\pm 1$ by rescaling $\lambda$, so that 
\begin{equation}
 \lambda-\lambda_{0}= \int^{\eta}_{\eta_{0}} a^{2}(\tilde{\eta})d\tilde{\eta},
\label{eq:affine_NG}
\end{equation}
where $\lambda=\lambda_{0}$ when $\eta=\eta_{0}$.
\item[(ii)] Timelike geodesic ($D=-1$). We again take $\dot{\eta}>0$, giving two cases. 
\begin{itemize}
 \item[(a)] Comoving timelike geodesic (CTG) ($C=0$): Equations~(\ref{eq:rdot}) and (\ref{eq:normalisation})
can be integrated to give
\begin{equation}
 \lambda-\lambda_{0}=\int^{\eta}_{\eta_{0}} a(\tilde{\eta}) d\tilde{\eta}=t-t_{0},\quad r=r_{0}.
\label{eq:affine_CTG}
\end{equation}
An observer on this 
world line sees 
the Universe isotropic.
 \item[(b)] Non-comoving timelike geodesic (NCTG) $(C\ne 0)$. Equations~(\ref{eq:rdot}) and (\ref{eq:normalisation}) 
can be  integrated to give
\begin{eqnarray}
 \lambda-\lambda_{0}&=& \int^{\eta}_{\eta_{0}}\frac{a(\tilde{\eta})d\tilde{\eta}}{\sqrt{1+(C/a(\tilde{\eta}))^{2}}} \, , \label{eq:affine_NCTG}\\
 r-r_{0}&=&\int^{\eta}_{\eta_{0}} \frac{C d\tilde{\eta}}{\sqrt{a^{2}(\tilde{\eta})+C^{2}}} \, .
\label{eq:r_NCTG}
\end{eqnarray}
The latter can be transformed to 
\begin{equation}
 (\eta-\eta_{0})-\sigma(r-r_{0})=\int^{\eta}_{\eta_{0}} \left(1-\frac{1}{\sqrt{1+(a(\tilde{\eta})/C)^{2}}}\right)d\tilde{\eta} \, ,
\label{eq:eta-r_NCTG}\\
\end{equation}
where $\sigma := \mathrm{sgn}(dr/d\eta)$.
\end{itemize}
\item[(iii)] Spacelike geodesic ($D=1$).  Equation~(\ref{eq:Vpotential}) implies 
$C\ne 0$, so $a(\eta)$ cannot be greater than $|C|$ along spacelike geodesics.
This gives two cases.
\begin{itemize}
 \item[(a)] Instantaneous spacelike geodesic (ISG). If we assume $\eta=\eta_{0}=\mbox{const}$,
equations~(\ref{eq:EL0}) and (\ref{eq:Vpotential})
 imply $a'(\eta_{0})=0$ and $C=\pm a(\eta_{0})$.
Equation~(\ref{eq:rdot}) 
can then be integrated to give
$\lambda-\lambda_{0}=C(r-r_{0})$. 
This geodesic is an orbit of a Killing vector of spatial translation 
in the isometry group associated with the homogeneity of the spacelike hypersurface.
 \item[(b)] Non-instantaneous spacelike geodesic (NISG). The geodesic equations 
can be integrated to give
\begin{eqnarray}
 \lambda-\lambda_{0}&=&
\sigma_{\eta} \int^{\eta}_{\eta_{0}}\frac{a(\tilde{\eta})d\tilde{\eta}}{\sqrt{(C/a(\tilde{\eta}))^{2}-1}}, \label{eq:affine_NISG}\\
 r-r_{0}&=&\sigma_{\eta} \int^{\eta}_{\eta_{0}} \frac{C}{\sqrt{C^{2}-a^{2}(\tilde{\eta})}}d\tilde{\eta},
\label{eq:r-eta_NISG}
\end{eqnarray}
where $\sigma_{\eta} := \mathrm{sgn}(\dot{\eta})$. The latter equation can be transformed to 
\begin{equation}
 (\eta-\eta_{0})-\sigma (r-r_{0})=-\int_{\eta_{0}}^{\eta}\left(\frac{1}{\sqrt{1-(a(\tilde{\eta})/C)^{2}}}-1\right)d\tilde{\eta},
\label{eq:eta-r_SG}
\end{equation}
where $\sigma=\mathrm{sgn}(dr/d\eta)$.
\end{itemize}
 \item[(iv)] Instantaneous spacelike curve (ISC). 
We consider a spacelike curve given by $\eta=\eta_{0}$. This is not a geodesic unless $a'(\eta_{0})=0$. The generalised affine parameter (g.a.p.) plays an important role in defining the 
b-boundary~\cite{Hawking:1973uf,Clarke:1994cw}
\footnote{The conformal boundary does not require the b-boundary construction. The b-boundary may be non-Hausdorff even for the FLRW 
models under certain conditions~\cite{Stahl:1999ix}.}.
If we choose the proper length $s=a r$ as a parameter along the curve, 
the tangent vector is 
$k^{a}=a^{-1}(\partial /\partial r)^{a}$, where $a=a(\eta_{0})$.
The p.p. basis $\{e^{a}_{(\alpha)}\}_{\alpha=0,1,2,3}$, which satisfies 
$k^{a}\nabla_{a}e_{(\alpha)}^{b}=0$, 
 is then given by 
\begin{eqnarray}
 e_{(0)a}&=& \cosh\left(\frac{a'}{a}r\right)a(d\eta)_{a}+\sinh \left(\frac{a'}{a}r\right)a(dr)_{a}, \nonumber \\
 e_{(1)a}&=& \sinh\left(\frac{a'}{a}r\right)a(d\eta)_{a}+\cosh \left(\frac{a'}{a}r\right)a(dr)_{a}, \nonumber \\
 e_{(2)a}&=& a\Sigma_{K}(r)(d\theta)_{a},\quad 
 e_{(3)a}=a\Sigma_{K}(r)\sin\theta (d\phi)_{a},
\end{eqnarray}
where $a=a(\eta_{0})$ and $a'=a'(\eta_{0})$.
Since 
\begin{equation}
 k^{(0)}=-\sinh\left(\frac{a'}{a}r\right),~k^{(1)}=\cosh\left(\frac{a'}{a}r\right),~k^{(2)}=k^{(3)}=0,
\end{equation}
where $k^{a}=k^{(\alpha)}e^{a}_{(\alpha)}$, 
the g.a.p. of this curve is given by the 
integral:
\begin{equation}
 u =\int_{0}^{s} \left[\sum_{\alpha=0}^{3}(k^{(\alpha)})^{2}\right]^{1/2}(\tilde{s}) d\tilde{s}=a \int_{0}^{r}\sqrt{\cosh\left(2\frac{a'}{a}\tilde{r}\right)}d\tilde{r}.
\end{equation}
Although this integral does not admit a compact expression 
in terms of elementary functions, 
its asymptotic form is
\begin{equation}
 u\approx \begin{cases}
	  ar & (2|a'|r/a\ll 1) \\
          \displaystyle\frac{a^{2}}{\sqrt{2}|a'|}e^{|a'|r/a} & (2|a'|r/a \gg 1) 
	 \end{cases}.
\end{equation}
\end{itemize}
Only null and comoving timelike geodesics were studied in Paper I but
here we extend the analysis to all geodesics and ISCs.

\subsection{Components of the Riemann tensor in the p.p. frame}

Since the FLRW spacetime is conformally flat, the Weyl tensor vanishes identically and the Riemann tensor is determined by the Ricci tensor, whose components 
in the coordinates $x^{\mu}=(\eta,r,\theta,\phi)$ are 
\begin{eqnarray}
 R_{00}&=&{\cal A}a^{2}, \quad {\cal A}a^{2}=-3{\cal H}', \\
 R_{0i}&=&0, \\
 R_{ij}&=& {\cal B}a^{2}\gamma_{ij}, \quad {\cal B}a^{2}={\cal H}'+2{\cal H}^{2}+2K,
\end{eqnarray}
where ${\cal H}:=a'/a$ and $i$ and $j$ run over 1, 2 and 3. 
Therefore, the Ricci tensor becomes
\begin{eqnarray}
 R_{ab}&=&{\cal A}a^{2}(d\eta)_{a} (d\eta)_{b}+{\cal B}a^{2}\gamma_{ij}(dx^{i})_{a} (dx^{j})_{b},
\end{eqnarray}
which can be rewritten as 
\begin{equation}
 R_{ab}={\cal A}e_{(0)a} e_{(0)b}+{\cal B}\delta^{ij}e_{(i)a} e_{(j)b},
\label{eq:Ricci_CTG_pp}
\end{equation}
where $\{e^{a}_{(\alpha)}\}_{\alpha=0,1,2,3}$ is 
a natural tetrad basis given by  
\begin{equation}
e_{(0)a}=-a(d\eta)_{a}, ~e_{(1)a}=a(dr)_{a}, ~e_{(2)a}=a\Sigma_{K}(r) (d\theta)_{a}, ~e_{(3)a}=a\Sigma_{K}(r)\sin\theta (d\phi)_{a}.
\label{eq:CTG_pp_basis}
\end{equation}
Since any scalar curvature polynomial constructed from the Riemann tensor 
can be written as a polynomial in ${\cal A}$ and ${\cal B}$, 
such as $R=-{\cal A}+3{\cal B}$ and $R^{ab}R_{ab}={\cal A}^{2}+3{\cal B}^{2}$, 
the boundedness of ${\cal A}$ and ${\cal B}$ implies that of 
all of the scalar curvature polynomials.

To examine the curvature singularities, we need to construct a p.p. basis 
along the pertinent curves. 
For CTGs, the p.p. orthonormal basis is given by 
equation~(\ref{eq:CTG_pp_basis})
and the Ricci tensor 
in the p.p. frame is given by equation~(\ref{eq:Ricci_CTG_pp}).
For null geodesics with
 tangent vector $k^{a}$, the p.p. pseudo-orthonormal basis is 
\begin{equation}
k_{a}=[-(d\eta)_{a}\pm (dr)_{a}]/\sqrt{2}, ~~l_{a}=a^{2}[-(d\eta)_{a}\mp (dr)_{a}]/\sqrt{2}, \label{eq:pp_basis_null} 
\end{equation}
where $e_{(2)a}$ and $e_{(3)a}$ are
as in equation~(\ref{eq:CTG_pp_basis}).
These 
satisfy
\begin{equation}
 k^{a}k_{a}=0, l^{a}l_{a}=0, k^{a}l_{a}=-1, k^{a}e_{(A)a}=0, l^{a}e_{(A)a}=0, g^{ab}e_{(A)a}e_{(B)b}=\delta_{AB},
\end{equation}
where $A$ and $B$ run over 2 and 3.
Using this basis, the Ricci tensor can be written as
\begin{equation}
R_{ab}=\frac{{\cal A}+{\cal B}}{2}a^{2}\left(k_{a}k_{b}
+\frac{1}{a^{4}}l_{a}l_{b}\right)+\frac{{\cal A}-{\cal B}}{2}(k_{a}l_{b}+l_{a}k_{b})
 +{\cal B}\delta^{AB}e_{(A)a}e_{(B)b}.
\label{eq:Ricci_null_pp}
\end{equation}
For general non-null geodesics with 
tangent vector $k^{a}$, the p.p. orthonormal basis is 
\begin{eqnarray}
e_{(0)a}=k_{a}=a^{2}[-k^{0}(d\eta)_{a}+k^{1}(dr)_{a}],~e_{(1)a}= a^{2}[-k^{1}(d\eta)_{a}+k^{0}(dr)_{a}],
\end{eqnarray}
where $e_{(2)a}$ and $e_{(3)a}$ are 
as in equation~(\ref{eq:CTG_pp_basis}).
In terms of this basis, the Ricci tensor becomes  
\begin{eqnarray}
 R_{ab}&=& \left[({\cal A}+{\cal B})\frac{C^{2}}{a^{2}}-{\cal A}D\right]e_{(0)a}e_{(0)b} 
\pm ({\cal A}+{\cal B})\frac{C}{a}\sqrt{\frac{C^{2}}{a^{2}}-D}(e_{(0)a}e_{(1)b}+e_{(1)a}e_{(0)b}) \nonumber \\
&& +\left[({\cal A}+{\cal B})\frac{C^{2}}{a^{2}}-{\cal B}D\right]e_{(1)a}e_{(1)b}
+{\cal B}\delta^{AB}e_{(A)a}e_{(B)b},
\label{eq:Ricci_pp_general_geodesic}
\end{eqnarray}
so the result for the CTGs is recovered 
if we take $C=0$ and $D=-1$.

The Einstein equation 
implies
\begin{equation}
{\cal A}=4\pi (\rho+3p),
\quad {\cal B}=4\pi (\rho-p).
\end{equation}
Thus, from the result in the previous section, an incomplete geodesic corresponds to an s.p. curvature singularity if and only if $\rho$ or $p$ is unbounded along the geodesic.
Using the conservation law, together with the equation of state $p=w\rho$, we find 
\begin{eqnarray}
{\cal A}=4\pi (1+3w)\rho_{0}\left(\frac{a_{0}}{a}\right)^{3(1+w)},
\quad {\cal B}=4\pi (1-w)\rho_{0}\left(\frac{a_{0}}{a}\right)^{3(1+w)}.
\end{eqnarray}

\section{Structure of conformal boundary}

\subsection{Definitions}

\subsubsection{Curvature singularity}
Although it is highly nontrivial to define spacetime singularities 
in general~\cite{Hawking:1973uf}, here we shall state that 
the spacetime is singular if it posseses at least one incomplete geodesic.
We define curvature singualrities below.
\begin{Def}
One has a curvature singularity 
if and only if a component of the Riemann tensor in the p.p. frame is unbounded 
along an incomplete geodesic.
This is also called a p.p. curvature singularity or 
${\bf C}^{0-}$ curvature 
singularity.
If and only if a scalar polynomial constructed 
from the Riemann tensor is unbounded along an incomplete geodesic, 
we call
it 
an s.p.  (scalar polynomial) curvature singularity.
An s.p. curvature singularity is a p.p. curvature singularity but not vice versa, 
so one can have
a non-scalar polynomial (non-s.p.) curvature singularity.
\end{Def}
The presence of a p.p. curvature singularity implies the inextendibility of the spacetime with a $C^{2-}$ 
metric
~\cite{Hawking:1973uf,Clarke:1994cw}, although it need not
 involve the divergence of fluid
quantities.
To take an example of non-s.p. curvature singularities, a pp-wave spacetime can admit a p.p. curvature singularity, while all the scalar curvature polynomials identically vanish~\cite{Stephani:2003tm}.
Here we consider 
the image of a curve in $M$ 
under the conformal transformation to the unphysical 
spacetime $\tilde{M}$.
Given an inextendible spacetime, 
if an incomplete geodesic 
terminates at a point on the conformal boundary of $M$, we call this ``endpoint'' 
a spacetime singularity.

\subsubsection{Conditions for p.p. curvature singularity}
Our discussion of the Riemann tensor 
in the p.p. frame for geodesics in the FLRW spacetime
implies the following. As for an incomplete CTG,  
equation~(\ref{eq:Ricci_CTG_pp}) implies that
the condition for a p.p. curvature singularity is the same as that for an s.p. curvature singularity.
From
equation~(\ref{eq:Ricci_null_pp}),  an incomplete null geodesic corresponds to a p.p. curvature singularity
if and only if at least one of
\begin{equation}
(\rho+p)a^{2}, ~(\rho+p)a^{-2}, ~\rho,
~ p 
\end{equation}
is unbounded, 
while from
equation~(\ref{eq:Ricci_pp_general_geodesic}) an incomplete non-null geodesic does so
if and only if at least one of
\begin{equation}
2(\rho+p)\frac{C^{2}}{a^{2}}-(\rho+3p)D, ~
(\rho+p)\frac{C}{a}\sqrt{\frac{C^{2}}{a^{2}}-D},~
2(\rho+p)\frac{C^{2}}{a^{2}}-(\rho-p)D,
~\rho-p 
\end{equation}
is unbounded.
Paper I only considered s.p. curvature singularities, while the current paper 
also includes p.p. curvature singularities.

\subsubsection{Conformal infinity and extendible boundary}
Before studying the structure of the conformal boundary for the FLRW solutions, 
we need further
definitions.
\begin{Def}
Let ${\mathscr S}$ be a subset of the conformal boundary.
If and only if any point in ${\mathscr S}$ is the endpoint of 
a future-directed complete null (timelike) geodesic in $M$, 
we call ${\mathscr S}$ a future null (timelike) 
infinity and denote it by ${\mathscr I}^{+}$ ($i^{+}$). 
A past null (timelike) infinity, denoted by ${\mathscr I}^{-}$ ($i^{-}$), is defined as its counterpart for 
a past-directed null (timelike) geodesic. 
If and only if 
any point in ${\mathscr S}$ 
is spacelike-related to all points 
in $M$ and 
also the endpoint of a b-complete spacelike curve in $M$, 
we call ${\mathscr S}$ a spacelike infinity and denote it by $i^{0}$.
So null (timelike) infinities are those for null (timelike) geodesics, while 
spacelike infinities are those for spacelike curves but not causal geodesics or curves.\footnote{
This terminology is
not the same as that in reference~\cite{Hawking:1973uf}, where 
infinities for null geodesics in de Sitter spacetime are termed 
spacelike infinities 
and denoted by ${\mathscr I}^{\pm}$.}
\end{Def}

\begin{Def}
Let ${\mathscr S}$ be a subset of the conformal boundary. 
If and only if any point in ${\mathscr S}$ 
is the endpoint of an incomplete geodesic but not a p.p. curvature singularity,
we call ${\mathscr S}$ an extendible boundary. 
\end{Def}

Although the above definitions may not suffice 
to classify the conformal boundaries of all possible spacetimes, 
they suffice for the classification 
of the FLRW spacetimes under consideration.
Note that if the spacetime $M$ is extendible, the above definition 
of spacelike infinities 
depends crucially on whether or not it is extended.
In the following, we restrict 
attention to the spacetime region described by the FLRW solutions 
discussed in Sec.~\ref{sec:FLRW_spacetime}, even if it is extendible.

\subsubsection{Big-bang, big-crunch and big-rip singularities}
\begin{Def}
For FLRW spacetimes, 
we call an s.p. curvature singularity at $t=t_{s}$ 
a big-bang (big-crunch) singularity if 
$a\to 0$ and $|\rho|\to \infty$ as $t\to t_{s}+ 0$ ($t\to t_{s}- 0$).
We call it 
a future (past) big-rip singularity if 
$a\to \infty $ and $|\rho|\to \infty$ 
as $t\to t_{s}- 0$ ($t\to t_{s}+ 0$).
\footnote{There was a typo in the corresponding definition in Paper I.} 
\end{Def}

\subsection{Flat FLRW solutions}
First we discuss the flat case
since this provides the basics
for other ones. 
The domains of $\eta$ are then 
$0<\eta<\infty$, $-\infty<\eta<\infty$ and $-\infty<\eta<0$ for 
$w>-1/3$, $w=-1/3$ and $w<-1/3$, respectively.
Geodesics in the Minkowski case
$(\rho=0)$ or the
$w\ge -1$ case with $\rho>0$ are well studied, e.g., in references~\cite{Hawking:1973uf,Senovilla:1998oua,Harada:2018ikn}. 
The results for the first
are summarised in table~
\ref{table:Minkowski_geodesics} and
those for the second  
 are shown as  F1, F2, F3 and dS in table~\ref{table:flat_geodesics}.
We 
 focus on $w<-1$ below and
some useful integrals are shown in Appendix~\ref{sec:integrals}.

\subsubsection{Null geodesics}
See the conformal completion diagram figure~\ref{fg:flat_domain}.
For null geodesics, as seen from equation~(\ref{eq:affine_NG}), 
the affine parameter $\lambda$ is determined by the integrals given by equation~(\ref{eq:flat_a2deta}).
For $w<-1$, in the limit $\eta\to 0$, which corresponds to the line segment
$\overline{\sf OE}$, $\lambda$ goes to $+\infty$ for $-5/3\le w<-1$, while 
it is finite for $w<-5/3$. Therefore, we identify $\overline{\sf OE}$ with ${\mathscr I}^{+}$ for $-5/3\le w<-1$ but 
with a big-rip singularity for $w<-5/3$.
As the past boundary $\eta=-\infty $ and $r=\infty$, which corresponds to 
$\overline{\sf ES}$, is approached, $\lambda$ is finite for $-5/3<w<-1$
but goes to $-\infty$ for $w\le -5/3$. 
For $-5/3<w<-1$, both $\rho$ and $p$ are bounded, 
while $(\rho+p)/a^{2}$ is unbounded, implying that $\overline{\sf ES}$ is identified with a non-s.p. curvature singularity.
For $w\le -5/3$, $\overline{\sf ES}$ is identified with ${\mathscr I}^{-}$.

\subsubsection{CTGs}
Next we consider CTGs. As seen from equation~(\ref{eq:affine_CTG}), 
$\lambda$ is determined by 
equation~(\ref{eq:flat_a1deta}).
For $w<-1$, the future boundary $\eta=0$, which corresponds to $\overline{\sf OE}$, 
is a big-rip singularity.
The past boundary $\eta=-\infty$, which corresponds 
to {\sf S},  is $i^{-}$.

\subsubsection{NCTGs}
For NCTGs, we first focus on the future boundary $\eta=0$.
As seen from equation~(\ref{eq:affine_NCTG}), 
$\lambda$ is determined by 
equation~(\ref{eq:flat_a1deta})
and it is finite for $w<-1$. 
From equations~(\ref{eq:r_NCTG}) and (\ref{eq:eta-r_NCTG}), 
both $r$ and $\eta-\sigma r$ remain finite in this limit.
Therefore,
from equation~(\ref{eq:tauchi_etar}),
the NCTGs terminate at $\overline{\sf OE}$ but not {\sf E}. 
Both $\rho$ and $p$ are unbounded, implying that $\overline{\sf OE}$ corresponds to 
a big-rip singularity.
Next, we consider the past boundary $\eta=-\infty$.
As seen from equation~(\ref{eq:affine_NCTG}), $\lambda$ 
is determined by 
equation~(\ref{eq:flat_a2deta}).
It is finite for $-5/3<w<-1$ but goes to $-\infty$ for $w\le -5/3$. From equation~(\ref{eq:eta-r_NCTG}), $\eta+r$ remains finite for $-5/3<w<-1$ but goes to $-\infty$ for $w\le -5/3$. For $-5/3<w<-1$, $\rho$ and $p$ are bounded but $(\rho+p)/a^{2}$ is unbounded. 
Therefore, 
from equation~(\ref{eq:tauchi_etar}),
for $-5/3<w<-1$, 
the NCTGs emanate from a non-s.p. curvature singularity on $\overline{\sf ES}$ but not from {\sf E} or {\sf S}, while for $w\le -5/3$ from {\sf S}, which is $i^{-}$. 

\subsubsection{Spacelike geodesics}
For spacelike geodesics, since $a'(\eta)\ne 0$, there is no ISG.
For $w<-1$, NISGs do not reach $\eta=0$ or $\overline{\sf OE}$.
For $\eta=-\infty$, $\lambda$ 
is determined by the integrals given by equation~(\ref{eq:flat_a2deta}).
Thus, it is finite for $-5/3< w<-1$ but goes to $+\infty$ for $w\le -5/3$.
From equations~(\ref{eq:eta-r_SG}) and (\ref{eq:flat_a2deta}), $\eta+r$ remains finite for $-5/3<w<-1$ but goes to $-\infty$ for $w\le -5/3$. For $-5/3<w<-1$, $\rho$ and $p$ are bounded but $(\rho+p)/a^{2}$ is unbounded. 
Therefore, 
from equation~(\ref{eq:tauchi_etar}),
for $-5/3<w<-1$, the NISGs emanate from or terminate at a non-s.p. curvature singularity on $\overline{\sf ES}$ but 
not {\sf E} or {\sf S}, while for $w\le -5/3$ they 
emanate from or terminate at {\sf E}, 
which is $i^{0}$. The coordinate $\eta$ cannot approach 
any nonzero finite value as $r\to \infty$.

\subsubsection{ISCs}
Finally, we consider ISCs. They are not geodesics in the flat case 
except for Minkowski spacetime.
They have infinite g.a.p. in the limit {\sf E}, which is $i^{0}$. 

\subsubsection{Summary of the flat solutions}
The results for all $K=0$ cases 
are summarised 
in tables~\ref{table:Minkowski_geodesics} and~\ref{table:flat_geodesics}.
Together with the conformal completion diagram figure~\ref{fg:flat_domain}, 
this implies the Penrose diagrams
shown in~figures~\ref{fg:Minkowski_penrose} and \ref{fg:flat_penrose}.
The latter corrects the 
misidentification of $\overline{\sf ES}$ with an extendible boundary in case F4a in figure 4 of Paper I.
 Care is needed for ${\mathscr I}^{+}$ if $a\to \infty$ 
and $-1<w<-1/3$.
In this case, since $\rho \propto a^{-3(1+w)}\to 0$ but $\rho a^{2}\propto a^{-(1+3w)}\to \infty$, a component of the Riemann tensor in the p.p. frame is unbounded in the infinite affine parameter limit. \footnote{This is a legitimate null infinity in a
classical spacetime. However, from a view point of quantum gravity, 
the curvature 
must exceed the Planck 
value at some finite 
$\lambda$,  so
the classical picture of spacetime 
must break down at this point.
This feature of ${\mathscr I}^{+}$ for $-1<w<-1/3$
does not depend on the spatial curvature.}
Care is also needed for $i^{0}$. Even for the standard case with $w>-1/3$, 
no spacelike geodesic emanates from or terminates at {\sf E} or $i^{0}$.
This is because $r$ is finite along a spacelike geodesic as
the integral on the right-hand side of equation~(\ref{eq:r-eta_NISG}) is finite if $|\eta|<\infty$.
This also applies
for $-5/3<w\le -1/3$ 
because $\eta+r$ is finite as the integral on the right-hand side of 
equation~(\ref{eq:eta-r_SG}) is 
finite in the limit $\eta\to -\infty$.

\begin{table}[H]
\begin{center}
\caption{Classification of geodesics in Minkowski spacetime. 
\label{table:Minkowski_geodesics}
}
\begin{tabular}{|c||c|c|c|c|c|c|}
\hline 
Case & --  & NG & CTG& NCTG & SG \\
\hline \hline
Minkowski & Vacuum & $\overline{\sf ES}$(${\mathscr I}^{-}$) $\to$ $\overline{\sf NE}$(${\mathscr I}^{+}$) & {\sf S}($i^{-}$) $\to$ {\sf N}($i^{+}$) & {\sf S}($i^{-}$) $\to$ {\sf N}($i^{+}$) & {\sf E}($i^{0}$)$\to$ {\sf E}($i^{0}$)\\ 
\hline
\end{tabular}  
\end{center}
\end{table}
~\footnote{
The points {\sf N}, {\sf S}, {\sf E} and {\sf O} are shown in 
figure~\ref{fg:flat_domain}.
For example, the expression `$\overline{\sf ES}$(${\mathscr I}^{-}$) $\to$ $\overline{\sf NE}$ (${\mathscr I}^{+}$)' in the $(2,1)$ cell means that null geodesics emanate from the line segment $\overline{\sf ES}$ and terminate at the line segment $\overline{\sf NE}$
and that $\overline{\sf ES}$ and $\overline{\sf NE}$ correspond to 
${\mathscr I}^{-}$ and ${\mathscr I}^{+}$,  
respectively.}

\begin{table}[H]
\begin{center}
\caption{Classification of geodesics in flat FLRW solutions. 
\label{table:flat_geodesics}
}
\begin{tabular}{|c||c|c|c|c|c|c|}
\hline 
Case & $w$ & NG & CTG& NCTG & SG \\
\hline \hline 
F1 & $(-1/3,\infty)$ & $\overline{\sf OE}$(sp) $\to$ $\overline{\sf NE}$(${\mathscr I}^{+}$) & $\overline{\sf OE}$(sp) $\to$ {\sf N}($i^{+}$) & $\overline{\sf OE}$(sp) $\to$ {\sf N}($i^{+}$) & $\overline{\sf OE}$(sp)$\to $$\overline{\sf OE}$(sp)\\
F2 & $-1/3$ & $\overline{\sf ES}$(sp) $\to$ $\overline{\sf NE}$(${\mathscr I}^{+}$) & {\sf S}(sp) $\to$ {\sf N}($i^{+}$) & $\overline{\sf ES}$(sp) $\to$ {\sf N}($i^{+}$) & $\overline{\sf ES}$(sp) $\to $ $\overline{\sf ES}$(sp) \\
F3 & $(-1,-1/3)$ & $\overline{\sf ES}$(sp) $\to$ $\overline{\sf OE}$(${\mathscr I}^{+}$) & {\sf S}(sp) $\to$ $\overline{\sf OE}$($i^{+}$) & $\overline{\sf ES}$(sp) $\to$ $\overline{\sf OE}$($i^{+}$) & $\overline{\sf ES}$(sp) $\to $ $\overline{\sf ES}$(sp) \\
dS & $-1$ & $\overline{\sf ES}$(ext) $\to$ $\overline{\sf OE}$(${\mathscr I}^{+}$) & {\sf S}($i^{-}$) $\to$ $\overline{\sf OE}$($i^{+}$) & $\overline{\sf ES}$(ext) $\to$ $\overline{\sf OE}$($i^{+}$) & $\overline{\sf ES}$(ext) $\to $ $\overline{\sf ES}$(ext) \\
F4a & $(-5/3,-1)$ & $\overline{\sf ES}$(nsp) $\to$ $\overline{\sf OE}$(${\mathscr I}^{+}$) & {\sf S}($i^{-}$) $\to$ $\overline{\sf OE}$(sp) & $\overline{\sf ES}$(nsp) $\to$ $\overline{\sf OE}$(sp) & $\overline{\sf ES}$(nsp) $\to $ $\overline{\sf ES}$(nsp)\\
F4b & $-5/3$ & $\overline{\sf ES}$(${\mathscr I}^{-}$) $\to$ $\overline{\sf OE}$(${\mathscr I}^{+}$) & {\sf S}($i^{-}$) $\to$ $\overline{\sf OE}$(sp) & {\sf S}($i^{-}$) $\to$ $\overline{\sf OE}$(sp) & {\sf E}($i^{0}$) $\to $ {\sf E}($i^{0}$) \\
F4c & $(-\infty,-5/3)$ & $\overline{\sf ES}$(${\mathscr I}^{-}$) $\to$ $\overline{\sf OE}$(sp) & {\sf S}($i^{-}$) $\to$ $\overline{\sf OE}$(sp) & {\sf S}($i^{-}$) $\to$ $\overline{\sf OE}$(sp) & {\sf E}($i^{0}$) $\to $ {\sf E}($i^{0}$) \\
\hline
\end{tabular} 
\end{center}
\end{table}
~\footnote{
The points {\sf N}, {\sf S}, {\sf E} and {\sf O} are shown in figure~\ref{fg:flat_domain}.
For example, the expression `$\overline{\sf OE}$(sp) $\to$ $\overline{\sf NE}$ (${\mathscr I}^{+}$)' in the $(2,1)$ cell means that for $w>-1/3$, null geodesics emanate from the line segment $\overline{\sf OE}$ and terminate at the line segment $\overline{\sf NE}$, 
with $\overline{\sf OE}$ and $\overline{\sf NE}$ corresponding to an s.p. curvature singularity and ${\mathscr I}^{+}$, respectively.
}

\subsection{Positive-Curvature FLRW solutions}

\subsubsection{Null and timelike geodesics}
For $K=1$, if $w\ne -1/3$, the solution is time-symmetric. The asymptotic behaviour of $a(\eta)$ at $\eta=0$ and $\eta=\eta_{c}=2\pi/(1+3w)$ coincides with that of flat FLRW at $\eta=0$. 
Therefore, the analysis of causal geodesics is straightforward.

\subsubsection{Spacelike geodesics for $w\ne -1/3$ \label{sec:P_spacelike_ne_-onethird}} 
The behaviour of spacelike geodesics is rather complicated. There exists an ISG at $\eta=\eta_{m}=\pi/(1+3w)$, which goes around the 3-sphere infinitely many times
for an infinitely large affine parameter. For $w>-1/3$, NISGs with $|C|>a(\eta_{m})$ 
emanate from the big-bang singularity at $\eta=0$ and terminate at the big-crunch singularity at $\eta=\eta_{c}$, NISGs with $|C|<a(\eta_{m})$ emanate from the big-bang singularity, bounce back and return to the big-bang singularity or behave like their time reverse with respect to $\eta=\eta_{m}$.  NISGs with $|C|=a(\eta_{m})$ emanate from the big-bang and approach
$\eta=\eta_{m}$, turning around the 3-sphere infinitely many times 
for an infinitely large affine parameter, or behave as their time reverse with respect to $\eta=\eta_{m}$. 
For $w<-1/3$, NISGs oscillate with respect to $\eta$ infinitely many times between $\eta_{+}$ and $\eta_{-}=\eta_{c}-\eta_{+}$, where $a(\eta_{\pm})=|C|$, and turn around the 3-sphere infinitely many times 
for an infinitely large affine parameter.

\subsubsection{Spacelike geodesics for $w=-1/3$ \label{sec:P_spacelike_-onethird}}
The $w=-1/3$ case requires special treatment. For $a= b_{0}e^{b_{c}\eta}$
($-\infty<\eta<\infty$), which we call case P2a, 
the analysis is essentially the same as for the flat solution with $w=-1/3$ 
except for the spatial spherical topology. 
There is no ISG, while NISGs emanate from and terminate at the big-bang singularity at $\eta=-\infty$.
For $a=a_{0}=\mathrm{const}$ ($-\infty<\eta<\infty$), which we call case P2b, the spacetime is given
by the exact Einstein static model.
In this case, all geodesics are complete.
There are ISGs at any $\eta$ and NISGs, both of which turn around the 3-sphere infinitely many times for an infinitely large affine parameter.

\subsubsection{Summary of the positive-curvature solutions}
The results for all cases are summarised in table~\ref{table:closed_geodesics}, except for spacelike geodesics because of their complicated behaviour.
Using figures~\ref{fg:closed_domain}, \ref{fg:closed_exception_domain} and \ref{fg:closed_domain_2}, we deduce the Penrose diagrams
shown in figure~\ref{fg:closed_penrose}.
This improves figure~5 of Paper I
by clarifying both $i^{\pm}$ and ${\mathscr I}^{\pm}$.

\begin{table}[H]
\begin{center}
\caption{Classification of causal geodesics in positive-curvature FLRW solutions. Cases P2a and P2b are an expanding solution and a static solution, respectively. See Secs.\ref{sec:P_spacelike_ne_-onethird} and \ref{sec:P_spacelike_-onethird} for spacelike geodesics.
\label{table:closed_geodesics}}
\begin{tabular}{|c||c|c|c|c|c|}
\hline 
Case & $w$ & NG & CTG& NCTG \\
\hline \hline
P1 & $(-1/3,\infty)$ & $\overline{\sf OE}$(sp) $\to$ $\overline{\sf NA}$(sp) & $\overline{\sf OE}$(sp) $\to$ $\overline{{\sf NA}}$(sp) & $\overline{\sf OE}$(sp) $\to$ $\overline{{\sf NA}}$(sp) \\
P2a & $-1/3$ & {\sf S}(sp) $\to$ {\sf N}(${\mathscr I}^{+}$) & {\sf S}(sp) $\to$ {\sf N}($i^{+}$) & {\sf S}(sp) $\to$ {\sf N}($i^{+}$) \\
P2b & $-1/3$ & {\sf S}(${\mathscr I}^{-}$) $\to$ {\sf N}(${\mathscr I}^{+}$) & {\sf S}($i^{-}$) $\to$ {\sf N}($i^{+}$) & {\sf S}($i^{-}$) $\to$ {\sf N}($i^{+}$) \\
P3 & $(-1,-1/3)$ & $\overline{\sf SB}$(${\mathscr I}^{-}$) $\to$ $\overline{\sf OE}$(${\mathscr I}^{+}$) & $\overline{\sf SB}$($i^{-}$) $\to$ $\overline{\sf OE}$($i^{+}$) & ${\overline{\sf SB}}$($i^{-}$) $\to$ $\overline{\sf OE}$($i^{+}$) \\
dS & $-1$ & ${\overline{\sf SB}}$(${\mathscr I}^{-}$) $\to$ $\overline{\sf OE}$(${\mathscr I}^{+}$) & ${\overline{\sf SB}}$($i^{-}$) $\to$ $\overline{\sf OE}$($i^{+}$) & ${\overline{\sf SB}}$($i^{-}$) $\to$ $\overline{\sf OE}$($i^{+}$) \\
P4a & $[-5/3,-1)$ & ${\overline{\sf SB}}$(${\mathscr I}^{-}$) $\to$ $\overline{\sf OE}$(${\mathscr I}^{+}$) & ${\overline{\sf SB}}$(sp) $\to$ $\overline{\sf OE}$(sp) & ${\overline{\sf SB}}$(sp) $\to$ $\overline{\sf OE}$(sp) \\
P4b & $(-\infty,-5/3)$ & ${\overline{\sf SB}}$(sp) $\to$ $\overline{\sf OE}$(sp)
& ${\overline{\sf SB}}$(sp) $\to$ $\overline{\sf OE}$(sp) & ${\overline{\sf SB}}$(sp) $\to$ $\overline{\sf OE}$(sp) \\
\hline
\end{tabular}  
\end{center}
\end{table}
~\footnote{The points 
{\sf N}, {\sf O}, {\sf E} and {\sf A} for case P1 are shown in figure~\ref{fg:closed_domain},
the points {\sf N}, {\sf S} and {\sf E} for cases P2a and P2b
in figure~\ref{fg:closed_exception_domain} and the points {\sf O}, {\sf S}, {\sf B} and {\sf E} for cases P3, dS, P4a and P4b 
in figure~\ref{fg:closed_domain_2}.
For example, the expression `$\overline{\sf OE}$(sp) $\to$ $\overline{\sf NA}$ (sp)' in the $(2,1)$ cell means that 
for $w>-1/3$, null geodesics emanate from the line segment $\overline{\sf OE}$ and terminate at the line segment $\overline{\sf NA}$,
with both $\overline{\sf OE}$ and $\overline{\sf NA}$ corresponding to s.p. curvature singularities.}

\subsection{Negative-curvature FLRW solutions with $\rho\ge 0$}

\subsubsection{Vacuum solution}

The vacuum solution is Milne spacetime, where there is no curvature singularity. 
Since $a=b_{0} e^{\eta}$, the 
behaviour of geodesics is the same as for the flat case with $w=-1/3$.
This behaviour is summarised in 
table~\ref{table:Milne_geodesics}
and figure~\ref{fg:open_domain} then gives
the Penrose diagram shown in figure~\ref{fg:Milne_penrose}.
\begin{table}[H]
\begin{center}
\caption{Classification of geodesics in Milne spacetime. 
\label{table:Milne_geodesics}}
\begin{tabular}{|c||c|c|c|c|c|c|}
\hline 
Case & --  & NG & CTG& NCTG & SG\\
\hline \hline
Milne & Vacuum & $\overline{\sf ES}$(ext) $\to$ $\overline{\sf NE}$(${\mathscr I}^{+}$) & {\sf S}(ext) $\to$ {\sf N}($i^{+}$) & $\overline{\sf ES}$(ext) $\to$ {\sf N}($i^{+}$) & $\overline{\sf ES}$(ext) $\to $ $\overline{\sf ES}$(ext) \\ 
\hline
\end{tabular}  
\end{center}
\end{table}
~\footnote{The points 
{\sf N}, {\sf S}, {\sf E} and {\sf O} are shown in figure~\ref{fg:open_domain}.
For example, the expression `$\overline{\sf ES}$(ext) $\to$ $\overline{\sf NE}$ (${\mathscr I}^{+}$)' in the $(2,1)$ cell means that null geodesics emanate from the line segment $\overline{\sf ES}$ and terminate at the line segment $\overline{\sf NE}$,
with $\overline{\sf ES}$ and $\overline{\sf NE}$ corresponding to an extendible boundary and ${\mathscr I}^{+}$, respectively.}

\subsubsection{Asymptotic behaviour of the non-vacuum solutions}
For $\rho>0$, we first consider $w\ne -1/3$. 
Since $a\propto \eta^{\alpha}$ in the limit $\eta \to 0$, the behaviour of geodesics is the same as in the flat case with the same $w$ in this limit.
The p.p.-frame 
components of the Riemann tensor also have the same behaviour as in the flat case.
In the limit $\tilde{\eta}\to  \infty$, since 
\begin{equation}
 \tilde{a}\approx \tilde{a}_{c}\frac{e^{\tilde{\eta}}}{4},\quad \eta=\frac{1}{1+3w}\tilde{\eta}
\end{equation}
and $\tilde{a}=a^{1+3w}$, we find 
\begin{equation}
 a\propto e^{\eta},
\end{equation}
as in the flat case with $w=-1/3$.

\subsubsection{Null, timelike and spacelike geodesics}
From the above discussion, if $w>-1/3$, in the limit $\eta\to \infty$, $a\to \infty$ and all causal geodesics are future complete.
All NCTGs terminate at {\sf N} in figure~\ref{fg:open_domain}.
For $w<-1/3$, in the limit $\eta\to -\infty$, $a\to 0$ and all causal geodesics are past incomplete. 
NCTGs emanate from $\overline{\sf ES}$ but 
not from {\sf E} or {\sf S}.
Since $\rho\propto a^{-3(1+w)}$ and $(\rho+p)/a^{2}\propto (1+w)a^{-(5+3w)}$,
$\overline{\sf ES}$ corresponds to an s.p. curvature singularity for $-1<w<-1/3$, 
a non-s.p. curvature singularity for $-5/3<w<-1$ 
and an extendible null boundary for $w\le -5/3$.
For $w=-1/3$, the behaviour of geodesics is the same as in the flat case with $w=-1/3$.
For $w>-5/3$, no spacelike geodesic emanates from or terminates at {\sf E} or $i^{0}$.

\subsubsection{Summary of the negative-curvature solutions with $\rho>0$}
The results for all negative-curvature FLRW solutions with $\rho> 0$ 
are summarised in 
table~\ref{table:open_positive_geodesics}.
Together with
figure~\ref{fg:open_domain}, this then gives
the Penrose diagrams
shown in figure~\ref{fg:open_positive_penrose}.
This corrects the misidentification of $\overline{\sf ES}$ with an extendible boundary for case NP4a in figure~7 of Paper I.
The present analysis shows that these solutions
are past extendible beyond the null boundary for $w\le -5/3$. Although the construction of the extended spacetime is beyond the scope of this paper, we infer that the collapsing solution obtained by 
time-reversing the expanding 
one could be pasted at the extendible null boundary, thereby 
 maximally extending the spacetime.

\begin{table}[H]
\begin{center}
\caption{Classification of geodesics in negative-curvature FLRW solutions with $\rho> 0$. 
\label{table:open_positive_geodesics}}
\begin{tabular}{|c||c|c|c|c|c|}
\hline 
Case & $w$ & NG & CTG& NCTG & SG \\
\hline \hline
NP1 & $(-1/3,\infty)$ & $\overline{\sf OE}$(sp) $\to$ $\overline{\sf NE}$(${\mathscr I}^{+}$) & $\overline{\sf OE}$(sp) $\to$ {\sf N}($i^{+}$) & $\overline{\sf OE}$(sp) $\to$ {\sf N}($i^{+}$) & $\overline{\sf OE}$(sp) $ \to $ $\overline{\sf OE}$(sp) \\
NP2 & $-1/3$ & $\overline{\sf ES}$(sp) $\to$ $\overline{\sf NE}$(${\mathscr I}^{+}$) & {\sf S}(sp) $\to$ {\sf N}($i^{+}$) & $\overline{\sf ES}$(sp) $\to$ {\sf N}($i^{+}$) & $\overline{\sf ES}$(sp) $\to$ $\overline{\sf ES}$ (sp) \\
NP3 & $(-1,-1/3)$ & $\overline{\sf ES}$(sp) $\to$ $\overline{\sf OE}$(${\mathscr I}^{+}$) & {\sf S}(sp) $\to$ $\overline{\sf OE}$($i^{+}$) & $\overline{\sf ES}$(sp) $\to$ $\overline{\sf OE}$($i^{+}$) & $\overline{\sf ES}$(sp) $\to $ $\overline{\sf ES}$(sp) \\
dS & $-1$ & $\overline{\sf ES}$(ext) $\to$ $\overline{\sf OE}$(${\mathscr I}^{+}$) & {\sf S}(ext) $\to$ $\overline{\sf OE}$($i^{+}$) & $\overline{\sf ES}$(ext) $\to$ $\overline{\sf OE}$($i^{+}$) & $\overline{\sf ES}$(ext) $ \to $ $\overline{\sf ES}$(ext) \\
NP4a & $(-5/3,-1)$ & $\overline{\sf ES}$(nsp) $\to$ $\overline{\sf OE}$(${\mathscr I}^{+}$) & {\sf S}(ext) $\to$ $\overline{\sf OE}$(sp) & $\overline{\sf ES}$(nsp) $\to$ $\overline{\sf OE}$(sp) & $\overline{\sf ES}$(nsp) $\to $ $\overline{\sf ES}$(nsp) \\
NP4b & $-5/3$ & $\overline{\sf ES}$(ext) $\to$ $\overline{\sf OE}$(${\mathscr I}^{+}$) & {\sf S}(ext) $\to$ $\overline{\sf OE}$(sp) & $\overline{\sf ES}$(ext) $\to$ $\overline{\sf OE}$(sp) & $\overline{\sf ES}$(ext) $\to$ $\overline{\sf ES}$(ext) \\
NP4c & $(-\infty,-5/3)$ & $\overline{\sf ES}$(ext) $\to$ $\overline{\sf OE}$(sp) & {\sf S}(ext) $\to$ $\overline{\sf OE}$(sp) & $\overline{\sf ES}$(ext) $\to$ $\overline{\sf OE}$(sp)& E($i^{0}$) $\to$ E($i^{0}$)\\
\hline
\end{tabular}  
\end{center}
\end{table}
~\footnote{The points 
{\sf N}, {\sf S}, {\sf E} and {\sf O} are shown in figure~\ref{fg:open_domain}.
For example, the expression `$\overline{\sf OE}$(sp) $\to$ $\overline{\sf NE}$ (${\mathscr I}^{+}$)' in the $(2,1)$ cell means that for $w>-1/3$, null geodesics emanate from the line segment $\overline{\sf OE}$ and terminate at the line segment $\overline{\sf NE}$,
with $\overline{\sf OE}$ and $\overline{\sf NE}$ corresponding to an s.p. curvature singularity and ${\mathscr I}^{+}$, respectively.}

\subsection{Negative-Curvature FLRW solutions with $\rho<0$}

\subsubsection{Null and timelike geodesics for $w>-1/3$}
We first consider $w\ne -1/3$. 
See figure~\ref{fg:open_domain}.
In this case, the solution is time-symmetric. 
In the limit $\tilde{\eta}\to \pm \infty$, since $\tilde{a}\propto e^{|\tilde{\eta}|}$, we have $a\propto e^{\mathrm{sgn}(1+3w)|\eta|}$.
For $w>-1/3$
and $\eta\to \pm \infty$, 
all causal geodesics are complete. 
Both CTGs and NCTGs emanate from {\sf S}, which is $i^{-}$, and terminate at {\sf N}, which is $i^{+}$.
The null boundaries $\overline{\sf ES}$ and $\overline{\sf NE}$ are 
${\mathscr I}^{-}$ and ${\mathscr I}^{+}$, respectively.

\subsubsection{Null and timelike geodesics for $w<-1/3$}
For $w<-1/3$
and $\eta\to \pm \infty$, all causal geodesics are incomplete. 
Since $\rho\propto a^{-3(1+w)}$, 
CTGs emanate from {\sf S}, a big-bang singularity, and terminate at {\sf N}, a big-crunch singularity, for $-1<w<-1/3$,  
but the spacetime is extendible beyond {\sf S} and {\sf N} for $w\le -1$.
Both null geodesics and NCTGs emanate from 
$\overline{\sf ES}$ but 
not from {\sf E} or {\sf S} and terminate at $\overline{\sf NE}$ but 
not at {\sf N} or {\sf E}.
Since $\rho\propto a^{-3(1+w)}$, $\rho a^{2}\propto a^{-(1+3w)}$ and $(\rho+p)/a^{2}\propto (1+w)a^{-(5+3w)}$, 
the null boundaries $\overline{\sf NE}$ and $\overline{\sf ES}$ are, respectively, big-bang and big-crunch singularities 
for $-1<w<-1/3$, null non-s.p. curvature singularities for $-5/3<w<-1$, 
and extendible boundaries for $w=-1$ and $w\le -5/3$.

\subsubsection{Spacelike geodesics for $w\ne -1/3$ \label{sec:NN_spacelike_ne_-onethird}}
The behaviour of spacelike geodesics in this case is rather complicated.
There exists an ISG at $\eta=0$, which 
emanates
from and terminates at {\sf E}, 
this being $i^{0}$. 
The behaviour of NISGs for $w>-1/3$ ($w<-1/3$) 
corresponds to that in the positive-curvature case for $w<-1/3$ ($w>-1/3$).
However, an open hyperbolic spatial geometry with future and past 
null boundaries 
replaces a 3-sphere with future and past 
spacelike boundaries.

\subsubsection{$w=-1/3$ \label{sec:NN_-onethird}}
Next we consider $w=-1/3$. For $0<\tilde{a}_{c}'<1$, which we call case NN2a,
the behaviour of the scale factor $a$ and the geodesics is
the same as in the flat case with $w=-1/3$.
There is no ISG, while NISGs emanate from and terminate 
at a big-bang null singularity $\overline{\sf ES}$.
For $\tilde{a}_{c}'=1$, which we call case NN2b, the spacetime is static and there is no curvature singularity.
All geodesics are complete. 
There are ISGs at any constant $\eta$ and all NISGs emanate from and terminate at 
{\sf E} or $i^{0}$.

\subsubsection{Summary for the negative-curvature solutions with $\rho<0$}
The results for all negative-curvature FLRW solutions with $\rho<0$ 
are summarised in table~\ref{table:open_negative_geodesics}
except for spacelike geodesics.
Figure~\ref{fg:open_domain} then gives 
the Penrose diagrams shown in figure~\ref{fg:open_negative_penrose}.
This corrects 
the misidentification of $\overline{\sf NE}$ and $\overline{\sf ES}$ with extendible boundaries in figure~8 of Paper I.
Although obtaining the extended spacetime is beyond the scope of the paper, 
we infer that the maximal extension for case NN4b ($w\le -5/3$)
may be obtained by 
pasting the same spacetime at the extendible null boundary infinitely many times.

\begin{table}[H]
\begin{center}
\caption{Classification of causal geodesics in negative-curvature FLRW solutions with $\rho<0$. Cases NN2a and NN2b are expanding and static solutions, respectively. See Secs.~\ref{sec:NN_spacelike_ne_-onethird} and 
\ref{sec:NN_-onethird} for spacelike geodesics. 
\label{table:open_negative_geodesics}}
\begin{tabular}{|c||c|c|c|c|c|}
\hline 
Case & $w$ & NG & CTG& NCTG \\
\hline \hline
NN1 & $(-1/3,\infty)$ & $\overline{\sf ES}$(${\mathscr I}^{-}$) $\to$ $\overline{\sf NE}$(${\mathscr I}^{+}$) & {\sf S}($i^{-}$) $\to$ {\sf N}($i^{+}$) & {\sf S}($i^{-}$) $\to$ {\sf N}($i^{+}$) \\
NN2a & $-1/3$ & $\overline{\sf ES}$(sp) $\to$ $\overline{\sf NE}$(${\mathscr I}^{+}$) & {\sf S}(sp) $\to$ {\sf N}($i^{+}$) & $\overline{\sf ES}$(sp) $\to$ {\sf N}($i^{+}$) \\
NN2b & $-1/3$ & $\overline{\sf ES}$(${\mathscr I}^{-}$) $\to$ $\overline{\sf NE}$(${\mathscr I}^{+}$) & {\sf S}($i^{-}$) $\to$ {\sf N}($i^{+}$) & {\sf S}($i^{-}$) $\to$ {\sf N}($i^{+}$) \\
NN3 & $(-1,-1/3)$ & $\overline{\sf ES}$(sp) $\to$ $\overline{\sf NE}$(sp) & {\sf S}(sp) $\to$ {\sf N}(sp) & $\overline{\sf ES}$(sp) $\to$ $\overline{\sf NE}$(sp) \\
AdS & $-1$ & $\overline{\sf ES}$(ext) $\to$ $\overline{\sf NE}$(ext) & {\sf S}(ext) $\to$ {\sf N}(ext) & $\overline{\sf ES}$(ext) $\to$ $\overline{\sf NE}$(ext) \\
NN4a & $(-5/3,-1)$ & $\overline{\sf ES}$(nsp) $\to$ $\overline{\sf NE}$(nsp) & {\sf S}(ext) $\to$ {\sf N}(ext) & $\overline{\sf ES}$(nsp) $\to$ $\overline{\sf NE}$(nsp) \\ 
NN4b & $(-\infty,-5/3]$ & $\overline{\sf ES}$(ext) $\to$ $\overline{\sf NE}$(ext) & {\sf S}(ext) $\to$ {\sf N}(ext) & $\overline{\sf ES}$(ext) $\to$ $\overline{\sf NE}$(ext)  \\
\hline
\end{tabular}  
\end{center}
\end{table}
~\footnote{
The points 
{\sf N}, {\sf S}, {\sf E} and {\sf O} are shown in figure~\ref{fg:open_domain}.
For example, the expression `$\overline{\sf ES}$(${\mathscr I}^{-}$) $\to$ $\overline{\sf NE}$ (${\mathscr I}^{+}$)' in the $(2,1)$ cell means that 
for $w>-1/3$,
null geodesics emanate from the line segment $\overline{\sf ES}$ and terminate at the line segment $\overline{\sf NE}$,
with $\overline{\sf ES}$ and $\overline{\sf NE}$ corresponding to ${\mathscr I}^{-}$ and ${\mathscr I}^{+}$, respectively.}

\acknowledgments
The authors are very grateful to T. Hiramatsu, D. Ida, A. Ishibashi, M. Kimura, H. Maeda, K.~Nomura, 
J.~M.~M. Senovilla, C.-M. Yoo and D. Yoshida for fruitful discussions. 
They are also grateful to anonymous referees for
constructive criticisms and comments. This work was partially supported by JSPS KAKENHI Grant Numbers JP19K03876, JP19H01895, JP20H05853 (TH) and JP19K14715 (TI).

\appendix  

\section{Abbreviations}
\label{sec:abbreviations}
The following abbreviations are used in this paper.\\
\\
\begin{tabular}{ll}
 AdS ~&~ Anti-de Sitter\\
CTG ~&~ Comoving timelike geodesic \\
dS  ~&~ de Sitter \\
 ext ~&~ extendible \\
 FLRW ~&~ Friedmann-Lema\^{i}tre-Robertson-Walker \\
 g.a.p. ~&~ generalised affine parameter\\
 ISC ~&~ Instantaneous spacelike curve \\
 ISG ~&~ Instantaneous spacelike geodesic \\
 NCTG ~&~ Non-comoving timelike geodesic    \\
 NG  ~&~ Null geodesic \\
 NISG ~&~ Non-instantaneous spacelike geodesic \\
 nsp, non-s.p.  ~&~ non-scalar polynomial \\
 SG   ~&~ Spacelike geodesic \\
 pp, p.p.   ~&~ parallelly propagated \\
 sp, s.p.   ~&~ scalar polynomial 
\end{tabular}

\section{Integrals}
\label{sec:integrals}
The following integrals are used in analysing geodesics for the flat FLRW solutions:
\begin{eqnarray}
 \int^{\eta}_{\eta_{0}}a(\tilde{\eta})d\tilde{\eta}&=& 
\begin{cases}
 \frac{1}{1+\alpha}b_{0}(\eta^{1+\alpha}-\eta_{0}^{1+\alpha}) & (w>-1/3) \\
 e^{b_{c}\eta}-e^{b_{c}\eta_{0}} & (w=-1/3) \\
 -\frac{1}{1+\alpha}b_{0}[(-\eta)^{1+\alpha}-(-\eta_{0})^{1+\alpha}] & (w<-1,-1<w<-1/3) \\
-b_{0}[\ln (-\eta)-\ln(-\eta_{0})] & (w=-1)
\end{cases}
, \label{eq:flat_a1deta}\\
 \int^{\eta}_{\eta_{0}}a^{2}(\tilde{\eta})d\tilde{\eta}&=&
\begin{cases}
 \frac{1}{1+2\alpha} b_{0}^{2}(\eta^{1+2\alpha}-\eta_{0}^{1+2\alpha}) & (w>-1/3) \\
 \frac{b_{c}}{2}(e^{2b_{c}\eta}-e^{2b_{c}\eta_{0}}) & (w=-1/3) \\
 -\frac{1}{1+2\alpha} b_{0}^{2}[(-\eta)^{1+2\alpha}-(-\eta_{0})^{1+2\alpha}] & (w<-5/3,-5/3<w<-1/3) \\
 -b_{0}^{2}[\ln (-\eta)-\ln(-\eta_{0})] & (w=-5/3)
\end{cases}
,
\label{eq:flat_a2deta}
\end{eqnarray}
where $\alpha=2/(1+3w)$.

\end{document}